\documentclass{ametsocV6.1}

\title{Disentangling North Atlantic ocean-atmosphere coupling using circulation analogues}

\authors{Matthew Patterson\aff{a}\correspondingauthor{Matthew Patterson, matthew.patterson@physics.ox.ac.uk}, Christopher O'Reilly \aff{b}, Jon Robson \aff{b,c}, Tim Woollings \aff{a}}

\affiliation{\aff{a}{Atmospheric, Oceanic and Planetary Physics, University of Oxford, UK}, \aff{b}{Department of Meteorology, University of Reading, UK}, \aff{c}{National Centre for Atmospheric Science, University of Reading, UK} }

\abstract{The coupled nature of the ocean-atmosphere system frequently makes understanding the direction of causality difficult in ocean-atmosphere interactions. This study presents a method to decompose turbulent heat fluxes into a component which is directly forced by atmospheric circulation, and a residual which is assumed to be primarily `ocean-forced'. This method is applied to the North Atlantic in a 500-year pre-industrial control run using the Met Office's HadGEM3-GC3.1-MM model. The method identifies residual heat flux modes largely associated with variations in ocean circulation and shows that these force equivalent barotropic circulation anomalies in the atmosphere. The first of these modes is characterised by the ocean warming the atmosphere along the Gulf Stream and North Atlantic Current and the second by a dipole of cooling in the western subtropical North Atantic and warming in the sub-polar North Atlantic. Analysis of atmosphere-only simulations confirms that these heat flux patterns are indeed forcing the atmospheric circulation changes seen in the pre-industrial control run. It is found that the Gulf Stream plays a critical role in the atmospheric circulation response to decadal ocean variability in this model. More generally, the heat flux dynamical decomposition method provides a useful way to establish causality in ocean-atmosphere interactions which could easily be applied to other ocean basins and to either models or reanalysis datasets.}

\begin{document}

\footnote{This Work has been submitted to the Journal of Climate. Copyright in this Work may be transferred without further notice.}

\maketitle

%
%
%
\statement
Variability of the ocean affects atmospheric circulation and provides a source of long-term predictability for surface weather. However, the atmosphere also affects the ocean. This makes separation of cause and effect in such atmosphere-ocean interactions difficult. This paper introduces a method to separate “turbulent heat fluxes”, the primary means by which the atmosphere and ocean influence one another, into a component driven by atmospheric variability and a component which is primarily related to ocean variability. The method is tested by applying it to a climate model simulation. The results suggest that the Gulf Stream region plays a strong role in driving the atmospheric response to ocean variability on decadal timescales.  A future paper will test the method using observational data. 

%
%
%

%


\section{Introduction}
Understanding the impact of extratropical sea surface temperature (SST) variability on atmospheric circulation is a complex problem. It is also an eminently practical one as ocean variability is a major source of prediction skill for forecasts on sub-seasonal to decadal timescales \citep{meehl_initialized_2021,merryfield_current_2020}. For example, on seasonal timescales, North Atlantic SST anomalies play some role in forcing the North Atlantic Oscillation \citep[NAO,][]{rodwell_oceanic_1999,mehta_oceanic_2000,czaja_observed_2002,gastineau_atmospheric_2013,dong_variability_2013,baker_linear_2019}. On longer timescales, Atlantic Multidecadal Variability \citep[AMV,][]{zhang_review_2019} has been linked to variability in the position of the Intertropical Convergence Zone (ITCZ) and Sahel rainfall \citep{knight_climate_2006}, multidecadal Atlantic hurricane activity \citep{sutton_climate_2007} and European climate \citep{sutton_atlantic_2012,oreilly_dynamical_2017}. 

Identifying the role of mid-latitude SSTs in atmospheric variability presents a challenge because of the coupled nature of the ocean-atmosphere system and the relatively weak influence of the ocean on the atmosphere \citep{kushnir_atmospheric_2002}. Turbulent heat exchange is the primary process by which the ocean and atmosphere transfer heat between each other \citep{gulev_north_2013}. On interannual and shorter timescales the atmosphere largely governs ocean-atmosphere covariability via modulation of air temperature, specific humidity and near-surface wind speed, and these in turn modify surface turbulent heat fluxes ($Q$, i.e. latent plus sensible heat fluxes). At decadal timescales and longer, the ocean dominates as it integrates atmospheric variability and responds via changes to ocean circulation and ocean heat transport, hence altering SSTs and $Q$ \citep{gulev_north_2013}. 

Idealised modelling studies have found that the initial response to a warm mid-latitude SST anomaly consists of a linear baroclinic response, with a downstream surface low advecting cool air equatorwards to balance the heating \citep{hoskins_steady_1981,hendon_stationary_1982}. After around ten days, the response becomes dominated by an equivalent barotropic pattern, involving transient eddy feedbacks, with the anomalous circulation extending far beyond the initial perturbation \citep{deser_transient_2007}. The equivalent barotropic response typically projects strongly onto the dominant modes of internal variability such as the NAO \citep{deser_effects_2004, peng_relationships_2001}. 

However, the circulation response to mid-latitude SSTs is substantially model dependent and may be sensitive to the location of the heating relative to the mean jet  \citep{ruggieri_atlantic_2021}. Modelled circulation responses to mid-latitude SSTs are also weak in comparison to both tropical SST-induced anomalies and internal, mid-latitude variability \citep{kushnir_atmospheric_2002}. However, the models' responses to mid-latitude SST variability are likely too weak \citep{eade_seasonal--decadal_2014,scaife_signal--noise_2018,smith_north_2020}, possibly due to low horizontal resolution \citep{scaife_does_2019} or weak eddy feedbacks \citep{hardiman_missing_2022}. Relatedly, models underestimate the magnitude of decadal to multidecadal extratropical variability of atmospheric circulation, particularly for the North Atlantic \citep{simpson_modeled_2018,oreilly_projections_2021}, even though many models overestimate decadal SST variability in the sub-polar North Atlantic \citep{patrizio_improved_2023-1}.



To identify SST-forcing of atmospheric circulation, studies use a variety of methods. These include 1) low-pass filtering data to isolate timescales at which the ocean dominates, 2) using lagged correlation analysis and 3) performing atmosphere-only experiments. However, as discussed above, current models are deficient at capturing the response to mid-latitude SSTs. Moreover, the shortness of the observed record means that low-pass filtering leaves only a few degrees of freedom, while lagged correlation analysis of a short time series can be hard to interpet.  

This study presents a method to separate the ocean-forced component of $Q$ from the atmospheric circulation-forced component. The method has been designed such that it does not require any low-pass filtering and can be applied to both models and reanalysis data. In this study the method is only applied to model data as this provides a more controlled setup in which there is a long data record and there is no observational uncertainty associated with variables such as $Q$. Testing with observation-based datasets, such as reanalysis, is reserved for a future paper. The method involves the use of circulation analogues to identify the component of $Q$ directly associated with circulation variability, diagnosing the ocean-forced component as the residual. We apply this method to a 500-year pre-industrial control run. We also test sensitivities of the results by applying it to simulations of the same model with observed external forcings from 1850 to 2014. 

The datasets and method are described in section \ref{Methods} before the method is applied to a piControl simulation in section \ref{PiControl}. The leading modes of the decomposition are examined in section \ref{Modes}, followed by an analysis of the circulation responses to the $Q$ modes in section \ref{Circulation} and an analysis of the differences in responses of the model at different resolutions to AMV in section \ref{Resolution}. Discussion and conclusions are provided in section \ref{Conclusions}. This is followed by appendices in which we examine the sensitivity to the choice of parameter values (appendix A) and sensitivity to the presence of external forcing and length of the dataset (appendix B).


\section{Data and methods}
\label{Methods}

\subsection{Data}
\label{Data}
We analyse simulations made using the UK Met Office HadGEM3-GC3.1 model \citep{williams_met_2018} for which North Atlantic ocean-atmosphere coupling has been extensively analysed \citep[e.g.][]{lai_mechanisms_2022,khatri_fast_2022}. The model consists of coupled ocean, atmosphere, land and sea-ice models. In this study, we primarily utilize the medium (MM) resolution version, but also briefly analyse the low (LL) resolution version. The `LL' and `MM' simulations are performed on N96 (grid-spacing of approximately 125km) and N216 (grid-spacing of approximately 60km) grids in the atmosphere, respectively. The horizontal ocean resolution is 0.25$^\circ$ (ORCA025) in `MM' and 1$^\circ$ (ORCA1) in `LL' but with a resolution of 0.33$^\circ$ from 15N to 15S. Both `LL' and `MM' have 75 vertical levels in the ocean and 85 in the atmosphere. We analyse a 500-year pre-industrial control simulation (piControl) using HadGEM3-GC3.1-MM and the first 500-years of a piControl simulation using HadGEM3-GC3.1-LL. These are fully coupled simulations which have no external forcing. Further details can be found in \citet{menary_preindustrial_2018} and \citet{kuhlbrodt_low-resolution_2018}. The piControl runs using both versions of the model simulate AMV with a 60-80 year period, consistent with observations, however the versions differ in terms of their ocean circulation variability and atmospheric response to AMV \citep{lai_mechanisms_2022}. Both versions show a slightly weaker Atlantic Meridional Overturning Circulation (AMOC) at 26.5N and at sub-polar latitudes with respect to RAPID \citep{menary_preindustrial_2018} and OSNAP observations \citep{menary_reconciling_2020}, respectively.

We also briefly analyse an atmosphere-only experiment, known as highresSST-present, taken from the HighResMIP project \citep{haarsma_high_2016}. In this experiment, HadGEM3-GC3.1-MM has been forced with observed SSTs \citep[taken from HadISST2,][]{titchner_met_2014} from 1950-2014 and with historical greenhouse gas and aerosol forcings. A total of three different ensemble members were run, all with slightly different initial states. Four members of historical coupled simulations from the same model \citep{andrews_historical_2020} are also analysed in appendix B. These use historical forcings including greenhouse gas and aerosol variations spanning 1850-2014. 



\subsection{Circulation analogues}
\label{Analogues}
In order to attribute $Q$ anomalies to atmospheric or oceanic forcing, we apply a circulation analogues method similar to that used by \citet{deser_forced_2016} and \citet{oreilly_dynamical_2017}. The concept of comparing similar circulation states was first developed in the context of statistical weather prediction by \citet{lorenz_atmospheric_1969} and later \citet{van_den_dool_searching_1994} and \citet{van_den_dool_performance_2003}. More recently, it has been used to study the degree to which atmospheric circulation trends have played a role in observed temperature trends \citep{cattiaux_winter_2010, wallace_simulated_2012, deser_forced_2016}.

The circulation analogues method attempts to estimate the component of a temporally and spatially varying variable, that is directly and simultaneously associated with changes in atmospheric circulation. In our case, we decompose $Q$ into two components, 

\begin{equation}
Q = Q_{CIRC} + Q_{RESIDUAL},
\end{equation}

 where $Q_{CIRC}$ is the atmospheric circulation-related component of $Q$ and $Q_{RESIDUAL}$, is the residual. We interpret the $Q_{RESIDUAL}$ to be primarily due to ocean variability and the persistence of SST anomalies forced by atmospheric circulation in previous months. $Q_{RESIDUAL}$ will also include small scale circulation features that do not project clearly on the monthly SLP fields. However, one can assume that these are random errors that should cancel out over large enough samples or over long enough periods. Additionally, $Q_{RESIDUAL}$ will vary due to radiative warming of the atmosphere, for example through externally forced radiative changes. There is no external forcing in the piControl run, but we remove the effects of external forcing when applying the method to historical simulations by regressing out global-mean SST anomalies from the SLP at each grid-point. This is discussed further in section \ref{Sensitivity}. We also perform a linear detrending of piControl anomalies prior to reconstructing SLP, in order to remove any drifts in the model.


Our method begins by taking the monthly-mean sea level pressure (SLP) anomalies for a particular month, say January 1901, and calculating the area-weighted Euclidian distance between this month and all other Januaries over the North Atlantic region (defined as 20$^{\circ}$N-75$^{\circ}$N, 90$^{\circ}$W-0$^{\circ}$E). Note that each month is only compared to its corresponding month (Januaries with Januaries etc), hence removing the seasonal cycle. A sub-sample of size $N_s$ from the $N_a$ most similar Januaries is then taken and a weighted sum of this sample of anomaly maps is used to reconstruct the original (January 1901) anomaly field. That is, we calculate values of weights, $\alpha_k$, which minimise the following

\begin{equation}
\sum_j \cos(\theta_j) \left |   SLP_{ij} - \sum_{k\neq i} \alpha_{k} SLP_{kj} \right | ,
\end{equation}

where $SLP_{ij}$ is the January SLP map for year $i$ and grid-point $j$ within the North Atlantic region. In this case, $i$ represents the year of the January which we are trying to fit to and $k$ is all other Januaries in the sub-sample. $\theta_j$ is the latitude of grid-point $j$, hence $\cos(\theta_j)$ is the area-weighting factor. Performing this minimisation gives a set of $N_s$ weights, $\alpha_k$, for each of the sub-sampled Januaries. $Q_{CIRC}$ is then calculated by summing the $Q$ anomalies for the same years multiplied by the corresponding weights calculated for the SLP anomalies, i.e.

\begin{equation}
Q_{CIRC_j} = \sum_k \alpha_k Q_{kj}.
\end{equation}

The resampling procedure is repeated $N_r$ times to obtain $N_r$ reconstructions of SLP and $Q_{CIRC}$, which are then averaged to find a best estimate of these two quantities. Here, $N_s$, $N_a$ and $N_r$ are taken to be 50, 80 and 100, respectively. This entire process is then repeated for all years in the dataset and for all calendar months of interest. For instance, if the December-January-February-March (DJFM) mean is required, the method is applied independently for December, January, February and March, before taking the average of these reconstructions.

\begin{figure}
  \centering
    \includegraphics[width=0.6\textwidth]{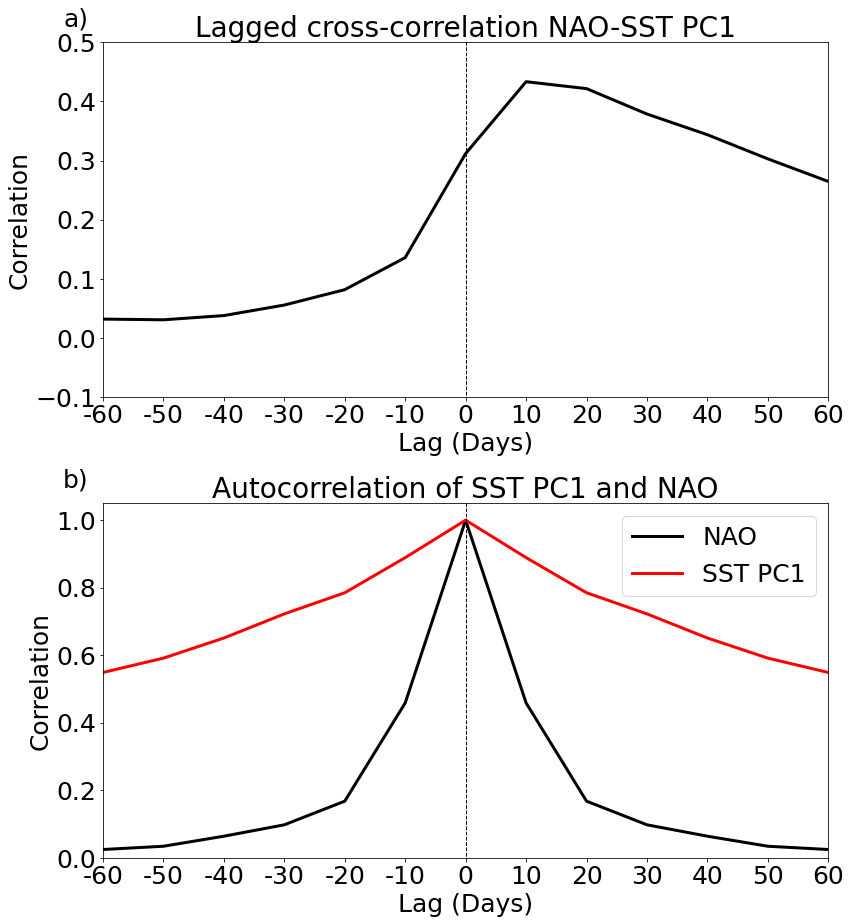}    
    \caption{a) Lagged cross-correlation of the NAO, calculated as the first principal component (PC) of SLP anomalies, 20$^{\circ}$N-80$^{\circ}$N, 60$^{\circ}$W-0$^{\circ}$E, with the first PC time series of SST, calculated over the same area for boreal winter (December-January-February-March). b) The autocorrelation of the NAO and SST PC time series'. The NAO is calculated using data from ERA5 \citep{hersbach_era5_2020} and the SST dataset is HadISST 2.1 \citep{titchner_met_2014}. The NAO leads the SSTs at positive lags and the SSTs lead at negative lags.}
\label{autocorrelation}
\end{figure} 

Applying the process separately to individual months before taking the seasonal average is a critical part of the procedure as this allows for the possibility of SSTs influencing atmospheric circulation. The point of the decomposition is to remove the direct influence that atmospheric circulation has on $Q$; for instance, through modulation of surface wind speed or the air-sea temperature difference. The correlation between atmospheric circulation and SSTs is strongest when the circulation leads by 10-20 days as the SSTs take some time to respond to the anomalous $Q$ \citep[figure \ref{autocorrelation}a,][]{deser_atmosphereocean_1997}. The autocorrelation timescale of the atmosphere is on the order of two weeks, in contrast to SST anomalies which persist for many months (figure \ref{autocorrelation}b). Consequently, SST anomalies created by stochastic atmospheric forcing in the early winter may then exert an influence on the atmosphere through to late winter. On the other hand, idealised experiments show that the atmospheric circulation response to mid-latitude SST anomalies takes several months to fully develop \citep{ferreira_transient_2005,deser_transient_2007}. Therefore, applying the method to monthly-mean data allows for circulation anomalies to develop in response to SST anomalies which were themselves induced by atmospheric forcing in prior months.  

The method is therefore not an attempt to completely separate the total influence of atmospheric circulation on $Q$, as the circulation may first influence SSTs, which then affect $Q$ for several months after as those SSTs persist. Rather, the dynamical decomposition is a diagnostic tool to measure the SST-driven component of $Q$ and consequently, establish how patterns of SST variability affect atmospheric circulation.

\subsection{Linear decomposition of $Q$ anomalies}
$Q$ is composed of sensible ($Q_S$) and latent heating ($Q_L$) terms which can be represented using the bulk formulae $Q_L = \rho C_e L U \Delta H$ and $Q_S = \rho C_p C_H U \Delta T$, respectively. Here, $\rho$ is the air density, $U$ is the near-surface wind speed, $C_p$ is the heat capacity of water, $L$ is the latent heat of evaporation and $C_e$ and $C_H$ are transfer coefficients. $\Delta H = H_s - H_a$ and $\Delta T = T_s - T_a$ are the air-sea temperature and specific humidity differences, respectively. Here subscript $s$ represents the sea surface and $a$, the atmosphere.

A linear decomposition of $Q$ \citep[e.g.][]{alexander_surface_1997,du_role_2008,he_north_2022} yields  

\begin{equation}
Q' =Q_S' + Q_L' \approx (\overline{Q_S} + \overline{Q_L}) \frac{U'}{\overline{U}} + \overline{Q_S}\frac{\Delta T'}{\overline{ \Delta T}} + \overline{Q_L}\frac{\Delta H'}{\overline{\Delta H}},
\end{equation}

where overbars represent time-mean quantities and primes are the anomalies with respect to the time-mean. This decomposition assumes that $U'\Delta H'<< \overline{U} \overline{\Delta H}$ and  $U'\Delta T'<< \overline{U} \overline{\Delta T}$, which are both good approximations at the monthly timescale \citep{alexander_surface_1997}. 

\subsection{Storm track diagnostics}
To analyse the atmospheric circulation response to SST variability, we calculate the transient meridional heat transport at 850hPa ($\overline{v'T'}$). Here, primes indicate band-pass filtering with a 10-day Lanczos filter \citep{duchon_lanczos_1979} and the overbar indicates a time-mean. We also utilize Hoskins E-vectors, defined as $\mathbf{E}=(\overline{v'^2-u'^2},\overline{-u'v'})$ \citep{hoskins_shape_1983}, where $u$ and $v$ are the zonal and meridional components of the wind, respectively. The divergence of the E-vectors indicates the acceleration of the mean flow by transient eddy momentum fluxes. 

\subsection{Indices}
The NAO index is calculated as the first empirical orthogonal function (EOF) of DJFM-mean SLP over the region 20$^{\circ}$N-80$^{\circ}$N and 60$^{\circ}$W-0$^{\circ}$W, calculated using the python package `eofs' \citep{dawson_eofs_2016}. The AMOC index is defined, following \citet{lai_mechanisms_2022}, as the Atlantic overturning stream function (in depth space) at 45$^{\circ}$N and 1000m depth. The AMV is calculated as the basin-mean North Atlantic SST (80$^{\circ}$W-0$^{\circ}$W, 0$^{\circ}$N-80$^{\circ}$N) after the global mean has been linearly removed from each grid-point. The index is then low-pass filtered using a 15-year running mean, again following \citet{lai_mechanisms_2022}.

\section{Evaluating the dynamical decomposition of $Q$ method}
\label{PiControl}

\subsection{Case study year}

\begin{figure}
  \centering
    \includegraphics[width=0.7\textwidth]{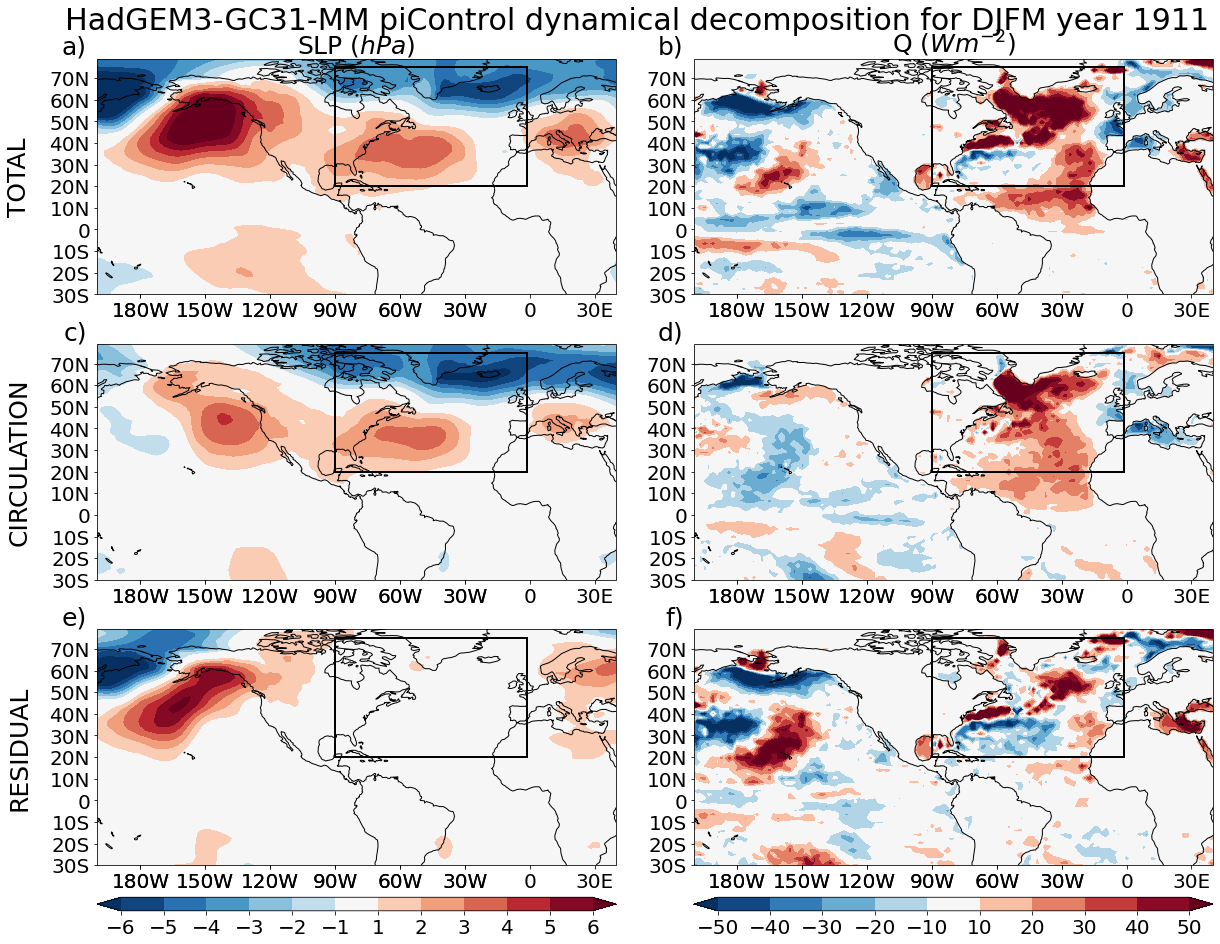}    
    \caption{Dynamical decomposition of a,c,e) SLP and b,d,f) $Q$ anomalies for the winter (DJFM)  of the year 1911 in the HadGEM3-GC31-MM piControl run. Shown are a,b) the full fields, c,d) the atmospheric circulation-related components and e,f) the residual components. The boxes in each panel indicate the North Atlantic region over which the decomposition is applied. Note that $Q$ is defined as positive from the ocean to the atmosphere.}
\label{reconstruction}
\end{figure} 

We now apply the circulation analogues method described in section \ref{Methods} to the North Atlantic, over a box bounded by latitudes 20$^{\circ}$N-75$^{\circ}$N and longitudes 90$^{\circ}$W-0$^{\circ}$E (shown by the boxed region in figure \ref{reconstruction}). An example of the decomposition is shown in figure \ref{reconstruction} for the winter (DJFM) of model year 1911 in the HadGEM3-GC31-MM piControl simulation. The circulation-related SLP field, marked by a postive-NAO like pattern is, by construction, almost identical to the full field over the North Atlantic region (figure \ref{reconstruction}a,c), however this is not the case outside of the North Atlantic (figure \ref{reconstruction}e). The $Q$ anomalies (defined as positive upwards throughout this study) indicate anomalously high heat loss from the ocean to the atmosphere over a horseshoe-shaped region involving the sub-polar North Atlantic and eastern sub-tropical North Atlantic (figure \ref{reconstruction}b). The dynamical decomposition suggests that a substantial proportion of this is related to the atmospheric circulation, including heat loss over the western sub-polar and subtropical North Atlantic (figure \ref{reconstruction}d). $Q_{RESIDUAL}$ anomalies are of similar magnitude to $Q_{CIRC}$ anomalies and are characterised by ocean heat-loss over the eastern sub-polar and heat-gain over the western subtropics, with a northward shift of the Gulf Stream (figure \ref{reconstruction}f). 

\begin{figure}
  \centering
    \includegraphics[width=\textwidth]{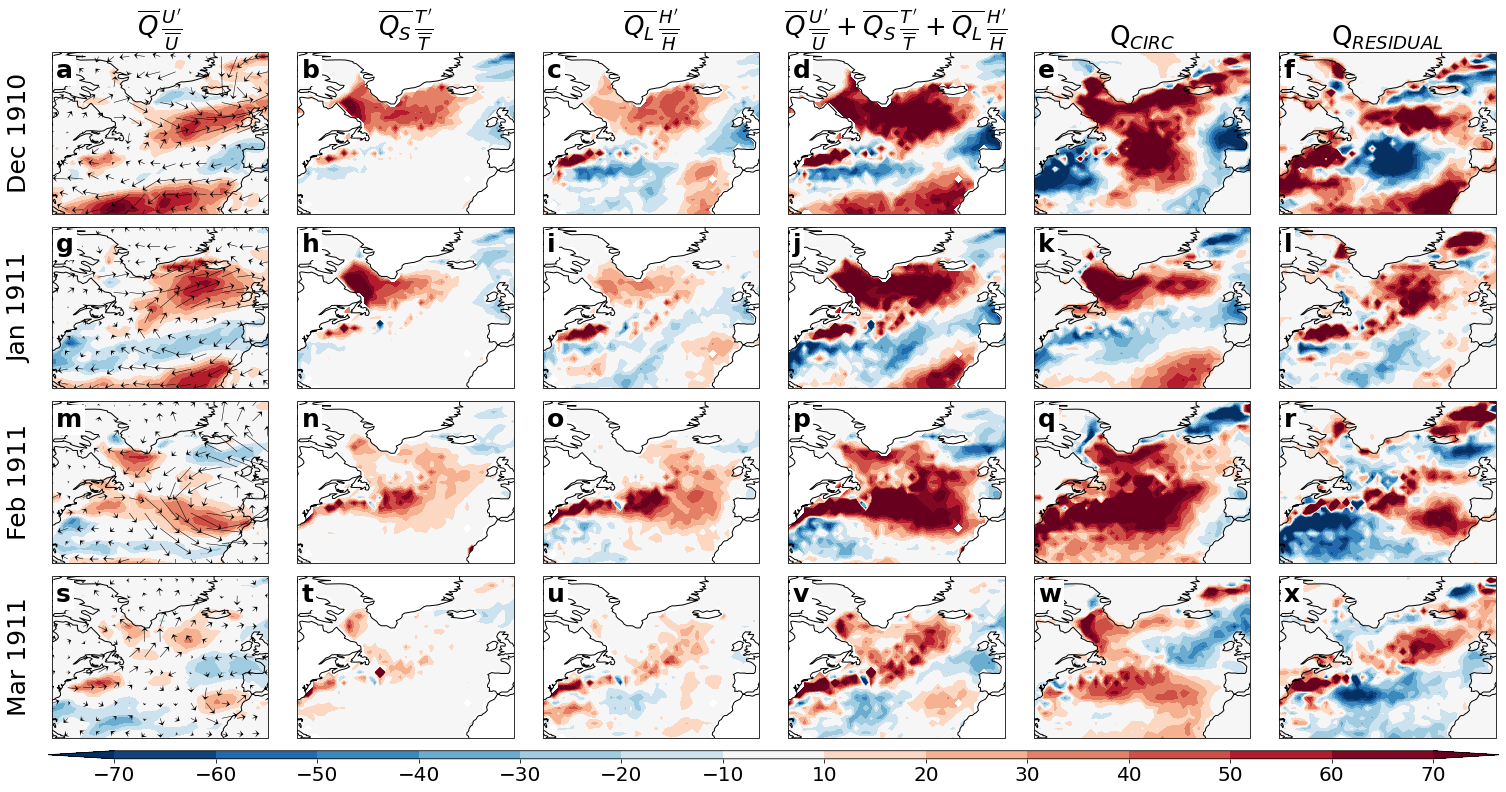}    
    \caption{Linear and dynamical decompositions of $Q$ anomalies for the winter of model year 1911 in the HadGEM3-GC3.1-MM piControl run. The different rows show results of the decompositions for a-f) December 1910, g-l) January 1911, m-r) February 1911 and s-x) March 1911. Columns show Q anomalies associated with a,g,m,s) surface-wind forcing, b,h,n,t) air-sea temperature differences, c,i,o,u) air-sea specific humidity differences, d,j,p,v) the sum of the first three columns, e,k,q,w) $Q_{CIRC}$ and f,l,r,x) $Q_{RESIDUAL}$. Vectors in a,g,m,s) show 10m wind anomalies. $Q$ anomalies are in units of $Wm^{-2}$.}
\label{decomposition}
\end{figure} 

The circulation analogues method is applied on a monthly basis and then averaged over DJFM. Therefore, in order to examine the contributions of different factors (wind speed, air-sea temperature and humidity differences) to $Q$ anomalies we need to look at individual months. Much of the sub-polar heat-loss over the sub-polar North Atlantic in winter 1911 occurs in December and January (compare figure \ref{decomposition}d,j with figure \ref{decomposition}p,v). Strong near-surface wind speeds and air-sea temperature and specific humidity differences all contribute to the sub-polar heat-loss (figure \ref{decomposition}a-c, g-i), which is largely driven by atmospheric circulation (figure \ref{decomposition}e,k). In all four months, $Q_{RESIDUAL}$ shows positive anomalies over the Gulf Stream and eastern sub-polar North Atlantic (figure \ref{decomposition}f,l,r,x) with negative anomalies in the subtropics from January to March (figure \ref{decomposition}l,r,x). The relative persistence of $Q_{RESIDUAL}$ compared to $Q_{CIRC}$ across the winter, is suggestive of the role of lower frequency SST variability in $Q_{RESIDUAL}$. Nevertheless, there are distinct $Q_{RESIDUAL}$ differences between each month, possibly due to atmospheric forcing from the previous month affecting SSTs.

\subsection{Interannual variability}

\begin{figure}
  \centering
    \includegraphics[width=\textwidth]{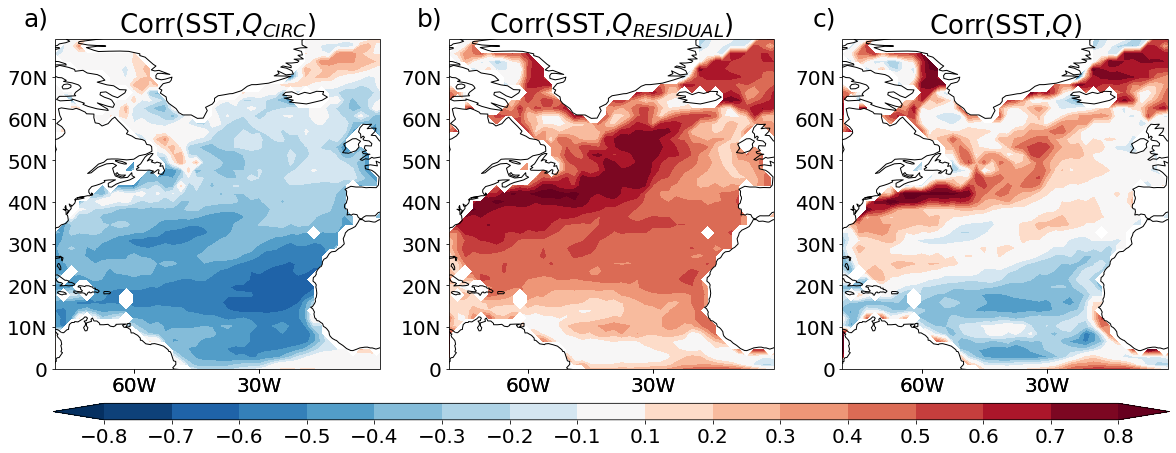}    
    \caption{Interannual (DJFM-mean) grid-point correlations between SST anomalies and a) $Q_{CIRC}$, b) $Q_{RESIDUAL}$ and c) $Q$ in the HadGEM3-GC1-MM piControl run.}
\label{correlation}
\end{figure}

\begin{figure}
  \centering
    \includegraphics[width=\textwidth]{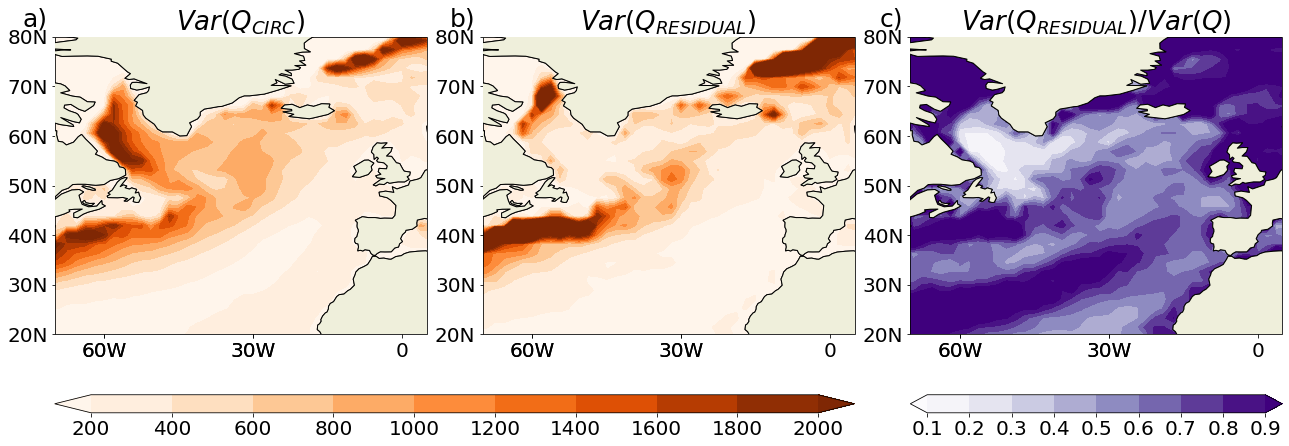}    
    \caption{Variance associated with interannual (DJFM) variability of the components of the $Q$ decomposition. Shown is the variance in a) $Q_{CIRC}$, b) $Q_{RESIDUAL}$ and c) the ratio of the variance of $Q_{RESIDUAL}$ to the variance in $Q$. Units for a,b) are both $(Wm^{-2})^2$ and c) is unitless.}
\label{variance}
\end{figure}

To test the dynamical decomposition method more systematically, we calculate the correlation, at each grid-point, between DJFM-mean SST anomalies and the components of the $Q$ dynamical decomposition. Consider that if a warm SST anomaly is primarily the result of warming by the atmosphere, then the anomalous $Q$ is negative, while a cool SST anomaly will be associated with a positive upward heat flux anomaly. Conversely, if the SST anomaly is warming or cooling the atmosphere, having formed through alterations to ocean circulation or by atmospheric forcing at least a month or two previous, then the sign of the SST anomaly should be the same as that of the anomalous heat flux. That is, a negative correlation between SST and $Q$ anomalies indicates a downward influence, while a positive correlation indicates an upward influence \citep[e.g.][]{gulev_north_2013,oreilly_signature_2016,bishop_scale_2017,Blackport_minimal_2019,oreilly_challenges_2023}. As anticipated, the $Q_{CIRC}$ component is negatively correlated with SST across the North Atlantic, suggesting a primarily downward influence (figure \ref{correlation}a). In contrast, SSTs are positively correlated with the $Q_{RESIDUAL}$ (figure \ref{correlation}b), suggesting a largely upward influence of $Q$ anomalies. For reference, the full $Q$ field shows that over the extratropical North Atlantic, SST variability tends to warm the atmosphere more than the atmosphere warms the SSTs, whereas the influence is generally downward in the tropical Atlantic (figure \ref{correlation}c). $Q$ variability in the Gulf stream region stands out as being particularly dominated by the ocean (figure \ref{correlation}c, figure \ref{variance}c). 

The majority of the simulated $Q$ variability in the sub-polar North Atlantic and particularly the Labrador Sea, is associated with $Q_{CIRC}$ (figure \ref{variance}a,c). This is likely due to atmospheric circulation modulating the advection of cold air from the North American continent over the ocean, raising the air-sea temperature contrast. The linear decomposition in figure \ref{decomposition} is consistent with this as $Q_S \Delta T'/\overline{\Delta T}$ in the Labrador Sea is larger than the other terms for all four months. $Q_{RESIDUAL}$ shows a similar magnitude of variability to $Q_{CIRC}$ along the North Atlantic Current (NAC) and larger variability associated with the Gulf Stream region. Therefore, while $Q_{CIRC}$ variability is larger over the sub-polar region, neither component of the decomposition completely dominates the $Q$ variability over the extratropical North Atlantic. The next section examines the primary modes of $Q$ variability associated with the components of the $Q$ decomposition and relates these to patterns of atmospheric circulation.

\section{Modes of $Q$ variability}
\label{Modes}

\begin{figure}
  \centering
    \includegraphics[width=\textwidth]{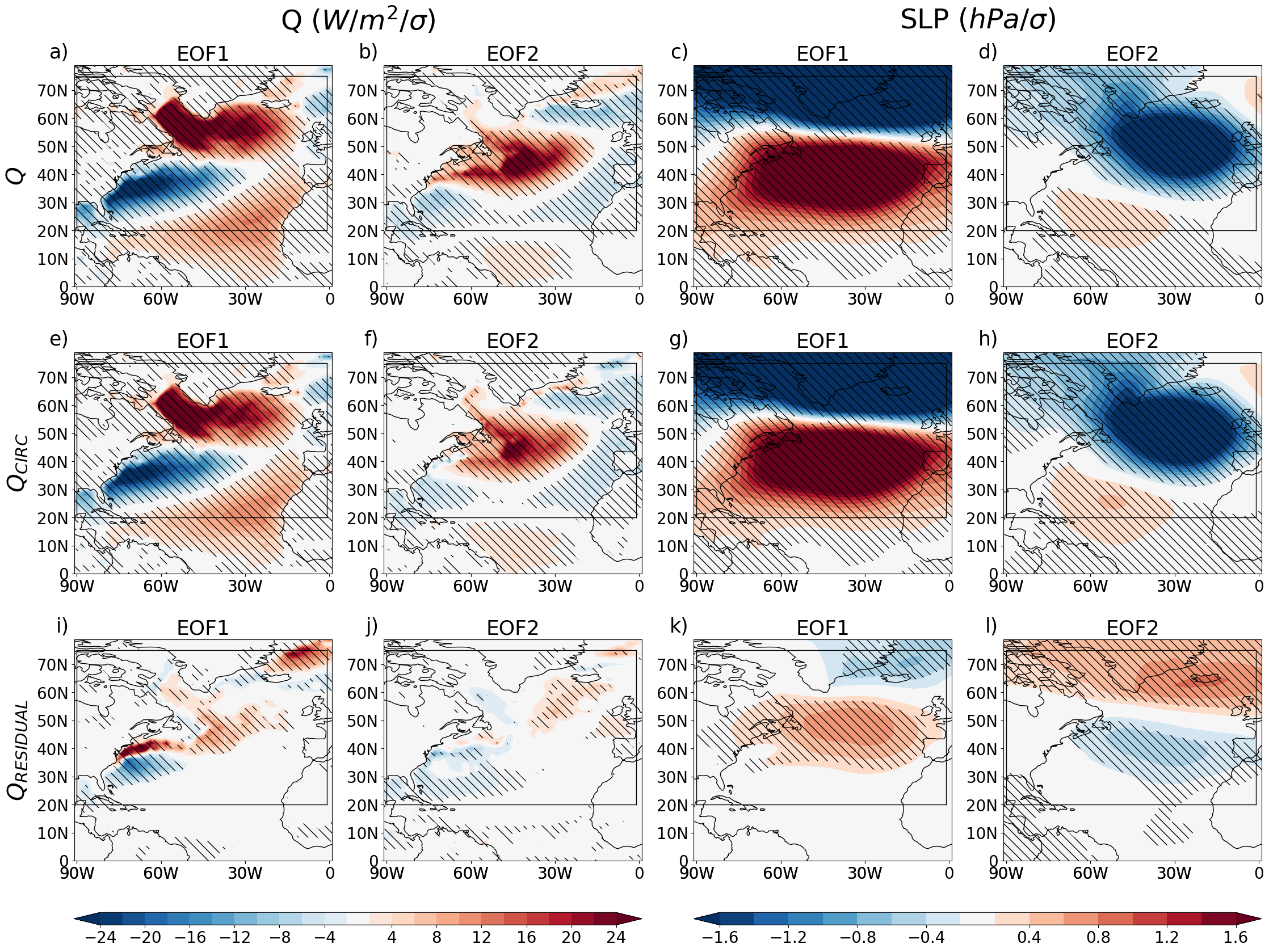}    
    \caption{Leading modes of variability associated with the components of the $Q$ decomposition. The first and second EOFs of a-d) $Q$, e-h) $Q_{CIRC}$ and i-l) $Q_{RESIDUAL}$ are shown regressed onto a-b,e-f,i-j) $Q$ and c-d,g-h,k-l) SLP. Hatching indicates where regression coefficients are statistically significantly different from zero, with p-values below 0.05, following a Student's t-test. The boxed region indicates the region over which both the decomposition and EOFs are calculated.}
\label{EOFs}
\end{figure} 

To understand the spatial patterns of variability associated with the $Q$ decomposition, we perform an area-weighted EOF analysis separately for $Q$, $Q_{CIRC}$ and $Q_{RESIDUAL}$, over the same region used to calculate the decomposition (20$^{\circ}$N-75$^{\circ}$N, 90$^{\circ}$W-0$^{\circ}$E). The first EOFs of $Q$ and $Q_{CIRC}$ are both characterised by a tripole pattern (figure \ref{EOFs}a,e), associated with the positive phase of the NAO (figure \ref{EOFs}c,g). The second EOFs of $Q$ and $Q_{CIRC}$ are again very similar to one another and consist of enhanced $Q$ over the central North Atlantic (figure \ref{EOFs}b,f), linked to the East Atlantic Pattern (figure \ref{EOFs}d,h). The first two EOFs of $Q_{CIRC}$ explain more variance (EOF1: 34.5\%, EOF2: 17.5\%) than those of $Q$ (EOF1: 25.0\%, EOF2: 12.6\%) likely because $Q$ also includes variability which is unrelated to atmospheric circulation. 

\begin{figure}
  \centering
    \includegraphics[width=\textwidth]{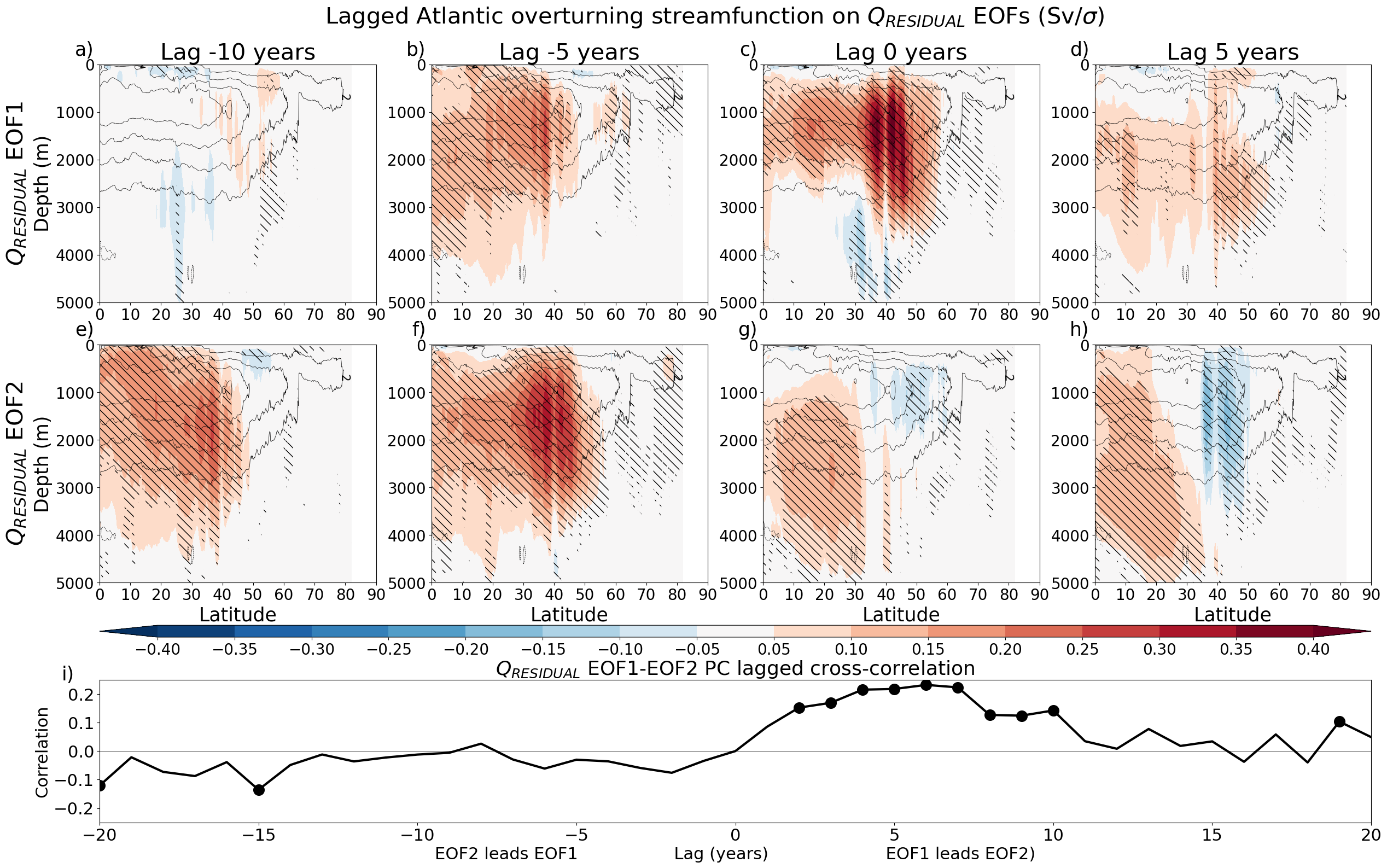}    
    \caption{Lagged regressions of the Atlantic overturning stream function (colours) onto $Q_{RESIDUAL}$ PC time series' associated with a-d) EOF1 and e-h) EOF2 as functions of depth and latitude. The mean overturning stream function is shown by unfilled contours with contours drawn every 4Sv, beginning at 2Sv. Hatching indicates statistical significance of regression coefficients, with p-values below 0.05, following a Student's t-test. i) Shows results of a lagged correlation analysis between $Q_{RESIDUAL}$ EOF1 and EOF2 PCs with the EOF1 time series leading at positive lags. Filled circles indicate statistically significant correlations, with p-values below 0.05, again following a Student's t-test.}
\label{AMOC}
\end{figure} 

The first two EOFs of $Q_{RESIDUAL}$ are more spatially localised. $Q_{RESIDUAL}$ EOF1 is marked by anomalous positive $Q$ along the NAC and Gulf Stream with a negative anomaly south of the Gulf Stream (figure \ref{EOFs}i). The second EOF has positive $Q$ anomalies to the south-east of Greenland, with a weaker negative anomaly in the subtropics (figure \ref{EOFs}j). Interestingly, the $Q_{RESIDUAL}$ EOFs are both associated with anomalous SLP patterns (figure \ref{EOFs}k,l). Given that the component of $Q$ related to simultaneous circulation variability has been removed from $Q_{RESIDUAL}$, we anticipate that these are indeed atmospheric circulation responses to the $Q_{RESIDUAL}$ patterns and will provide further evidence that this is the case in section \ref{Circulation}.

Returning to the $Q_{RESIDUAL}$ patterns, $Q_{RESIDUAL}$ EOF1 is somewhat reminiscent of the `slow' response at 3-4 year lags to NAO forcing found by \citet{khatri_fast_2022} using the same model but with decadal hindcast data (c.f. their figure 2). They found that the initial `fast' response to the NAO caused by wind stress and $Q$ anomalies is followed by a slower adjustment to SSTs involving a strengthened overturning circulation. To investigate the relationship of  $Q_{RESIDUAL}$ EOF1 with ocean overturning, we regress the Atlantic Overturning stream function onto the prinicpal component (PC) time series associated with $Q_{RESIDUAL}$ EOF1 in figure \ref{AMOC} at various lags. This analysis confirms that $Q_{RESIDUAL}$ EOF1 is associated with a strengthened AMOC (figure \ref{AMOC}c). The AMOC appears to strengthen about five years prior to the peak of $Q_{RESIDUAL}$ EOF1 (figures \ref{AMOC}b) and weakens following the $Q_{RESIDUAL}$ EOF1 peak (figures \ref{AMOC}d). $Q_{RESIDUAL}$ EOF2 is also linked to AMOC variability, though the peak AMOC strengthening occurs a few years prior to the $Q_{RESIDUAL}$ EOF2 peak (figures \ref{AMOC}f). The timing of the AMOC strengthening relative to the $Q_{RESIDUAL}$ EOF2 peak may relate to the fact that the $Q_{RESIDUAL}$ EOF1 and $Q_{RESIDUAL}$ EOF2 PCs are significantly, positively correlated when $Q_{RESIDUAL}$ EOF1 leads by 3-10 years. 

\begin{figure}
  \centering
    \includegraphics[width=\textwidth]{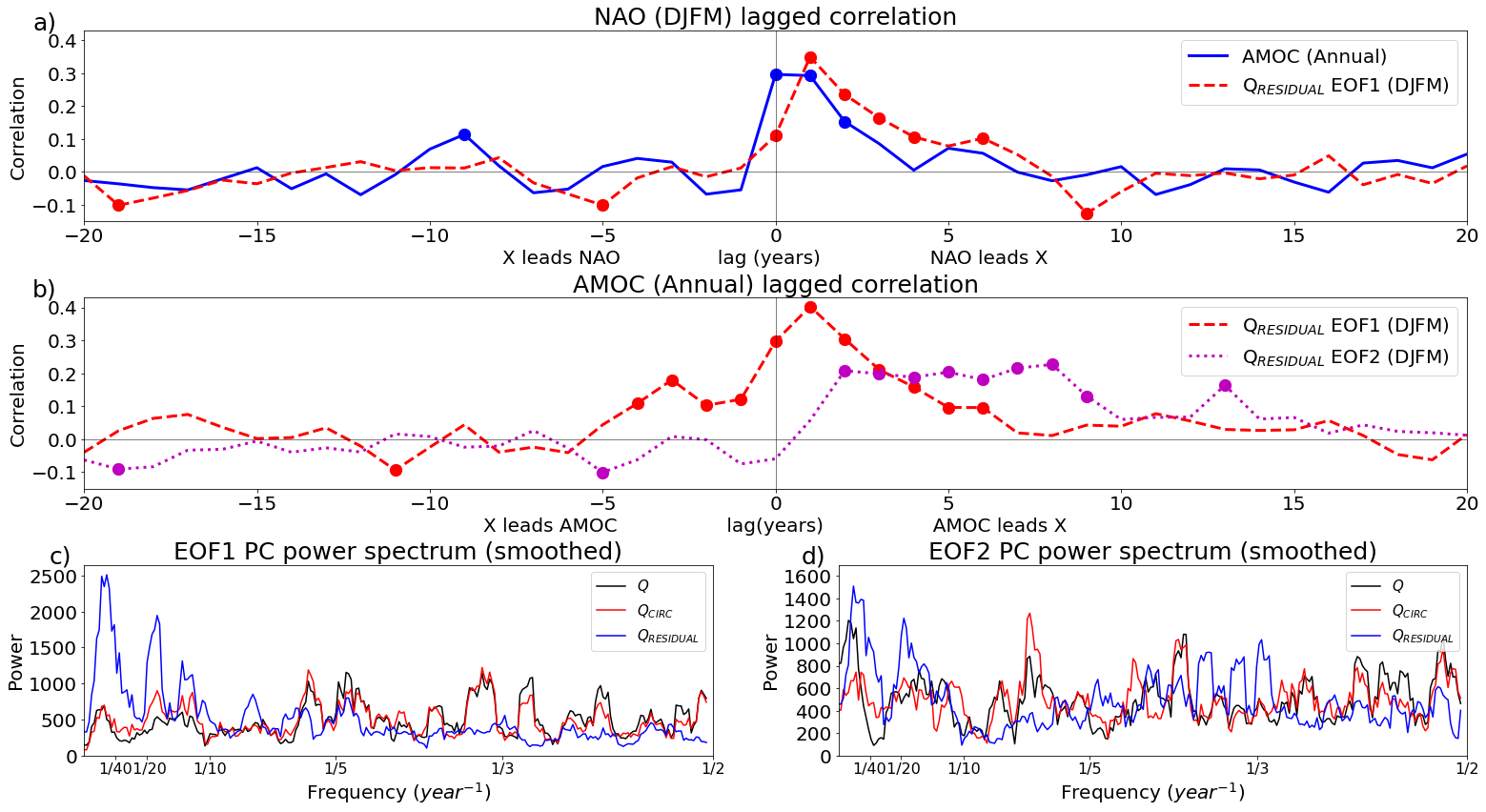}    
    \caption{Lagged correlations for a) the NAO index with the AMOC and $Q_{RESIDUAL}$ EOF1 PC time series, b) the AMOC with  $Q_{RESIDUAL}$ EOF1 and EOF2 PCs. Filled circles indicate statistically significant correlations with p-values below 0.05, following a t-test. Power spectra are also shown for the PCs associated with c) EOF1 and d) EOF2 of $Q$ (black), $Q_{CIRC}$ (red) and $Q_{RESIDUAL}$ (blue).}
\label{AMOC_corr}
\end{figure} 

The AMOC variability itself is partly driven by the NAO with slower ocean overturning integrating NAO variations over multiple years  \citep[e.g.][]{mccarthy_ocean_2015, oreilly_assessing_2019}. This can be seen in figure \ref{AMOC_corr}a, which shows the (annual-mean) AMOC is strengthened in the 2-3 years following a positive winter NAO. $Q_{RESIDUAL}$ EOF1 is most positively correlated with NAO variability when the NAO leads by 1 year, suggesting that the NAO strengthens the AMOC  (figure \ref{AMOC_corr}a), which subsequently alters $Q$ (figure \ref{AMOC_corr}b).  The signature of lower frequency ocean variability in $Q_{RESIDUAL}$ is also seen in power spectra of the $Q_{RESIDUAL}$ PCs. $Q_{RESIDUAL}$ EOF1 shows notable peaks in power for periods of 20 and 40 years, neither of which is seen for $Q$ and $Q_{CIRC}$ (figure \ref{AMOC_corr}c). $Q_{RESIDUAL}$ EOF2 shows peaks at similar frequencies (figure \ref{AMOC_corr}d). 

\begin{figure}
  \centering
    \includegraphics[width=\textwidth]{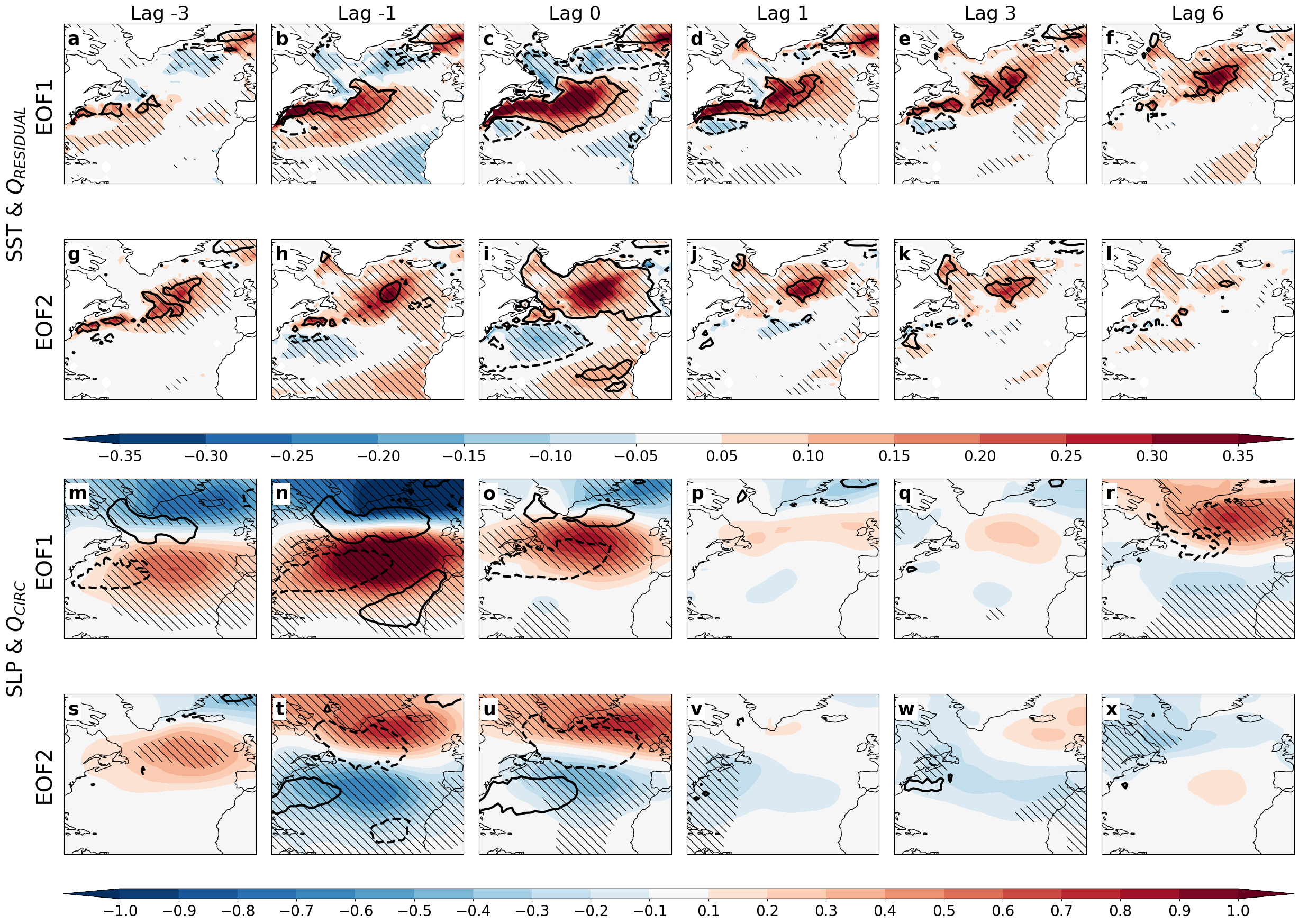}    
    \caption{Lagged regression with $Q_{RESIDUAL}$ EOF1 and EOF2 with SST and SLP in colours, and $Q_{RESIDUAL}$ and $Q_{CIRC}$ shown by unfilled contours. For $Q_{RESIDUAL}$ and $Q_{CIRC}$, contours are only drawn for $\pm3 Wm^{-2}$ for visual clarity. Hatching indicates where SST and SLP regression coefficients are statistically significant, with p-values below 0.05, following a Student's t-test.}
\label{lag}
\end{figure} 

Figure \ref{lag} shows lagged composites of the $Q_{RESIDUAL}$ EOFs and summarises the sequence of events preceding and following their peak. Positive $Q_{RESIDUAL}$EOF1 events are preceded by positive NAO forcing in the previous years which cools the western sub-polar North Atlantic and warms the Gulf Stream region (as indicated by $Q_{CIRC}$, the unfilled contours in figure \ref{lag}m-o). These changes drive a stronger AMOC and warm SSTs along the NAC (figure \ref{lag}a-c). These SST anomalies force an Atlantic ridge response (figure \ref{lag}o), which is investigated in the next section. The SST anomalies subsequently propagate towards the eastern sub-polar North Atlantic on timescales of 4-6 years (figure \ref{lag}c-f) until the pattern resembles $Q_{RESIDUAL}$ EOF2. The timescale for propagation of these anomalies is roughly consistent with observations \citep{arthun_anomalous_2016,arthun_skillful_2017} and analysis of an earlier version of HadGEM3 \citep{menary_mechanism_2015}. The North Atlantic is now marked by warm eastern sub-polar and cool Gulf Stream regions (figure \ref{lag}h-i). This subsequently appears to force a negative NAO (figure \ref{lag}r,t-u), which is also investigated in the next section. 

\section{Circulation response to heat flux anomalies}
\label{Circulation}

\begin{figure}
  \centering
    \includegraphics[width=\textwidth]{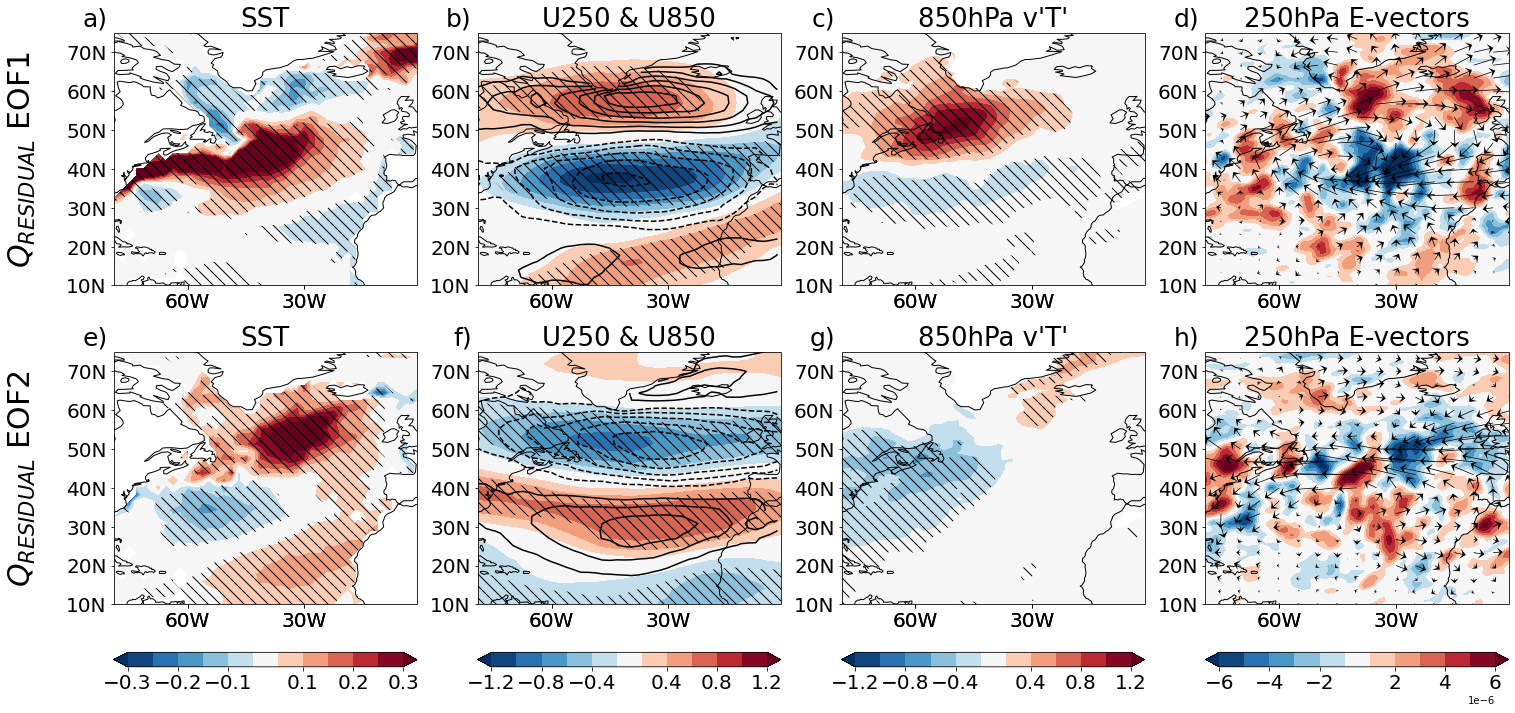}    
    \caption{Diagnostics showing the circulation response to $Q_{RESIDUAL}$ a-d) EOF1 and e-h) EOF2. Shown are regressions of a,e) SSTs, b,f) 250hPa (colours) and 850hPa (unfilled contours) winds, c,g) 850hPa 10-day high pass filtered meridional heat transport at 850hPa, d,h) E vectors (vectors) onto $Q_{RESIDUAL}$ EOF1 and EOF2 PCs. Colours in d,h) indicate the divergence associated with those vectors. Unfilled contours in b,f) are contoured every $0.1ms^{-1}/\sigma$. Hatching indicates statistical significance of regression coefficients with p-values less than 0.05, following a Student's t-test. }
\label{response}
\end{figure} 

\subsection{Storm track diagnostics}
Both $Q_{RESIDUAL}$ EOF1 and EOF2 are correlated with SLP anomalies at zero lag, suggesting that these are a response to the anomalous $Q$ as the direct circulation-forced component of $Q$ has been removed from $Q_{RESIDUAL}$ by construction. $Q_{RESIDUAL}$ EOF1, which is associated with positive SSTs along the NAC (figure \ref{response}a), forces a ridge between 40$^{\circ}$N and 60$^{\circ}$N, with an opposite-signed anomaly east of Greenland (figure \ref{EOFs}). This is also linked to an equivalent-barotropic northward shift of the jet and strengthening of storm track activity in the western North Atlantic, measured using transient heat transport, $\overline{v'T'}$, as a proxy (figure \ref{response}b,c). The increase in storm track activity is likely driven by the increased SST gradient along the northern flank of the Gulf Stream (figure \ref{response}a) and hence increased baroclinicity \citep{gan_feedbacks_2015,joyce_meridional_2019}. The paradigm of \citet{novak_life_2015} suggests that periods of high transient heat transport are associated with a higher frequency of downstream wave-breaking on the southward side of the jet. This decelerates the jet on the equatorward flank and transfers momentum poleward, deflecting the jet polewards. This is indicated by the divergence of E-vectors in figure \ref{response}d, which measure the acceleration of the mean flow by transient eddy momentum fluxes. The circulation response to the $Q_{RESIDUAL}$ EOF1 pattern, with high pressure over the North Atlantic, is similar to previous findings on the circulation response to Gulf Stream variability \citep[e.g.][]{ciasto_north_2004, wills_observed_2016, paolini_atmospheric_2022}.

On the other hand, $Q_{RESIDUAL}$ EOF2 appears to force a negative NAO-like pattern (figure \ref{EOFs}) and an equatorward shift of the eddy-driven jet. Understanding the response to EOF2 is complicated by the presence of two centres of action in the $Q_{RESIDUAL}$ pattern - one over the sub-polar North Atlantic and another in the Gulf Stream region (figure \ref{EOFs}j). The transient heat transport response for EOF2 is slightly weaker than for EOF1, but it is possible that the cooler Gulf Stream in EOF2 acts to shift the jet southward via a similar mechanism to EOF1 (but with signs reversed).


\begin{figure}
  \centering
    \includegraphics[width=\textwidth]{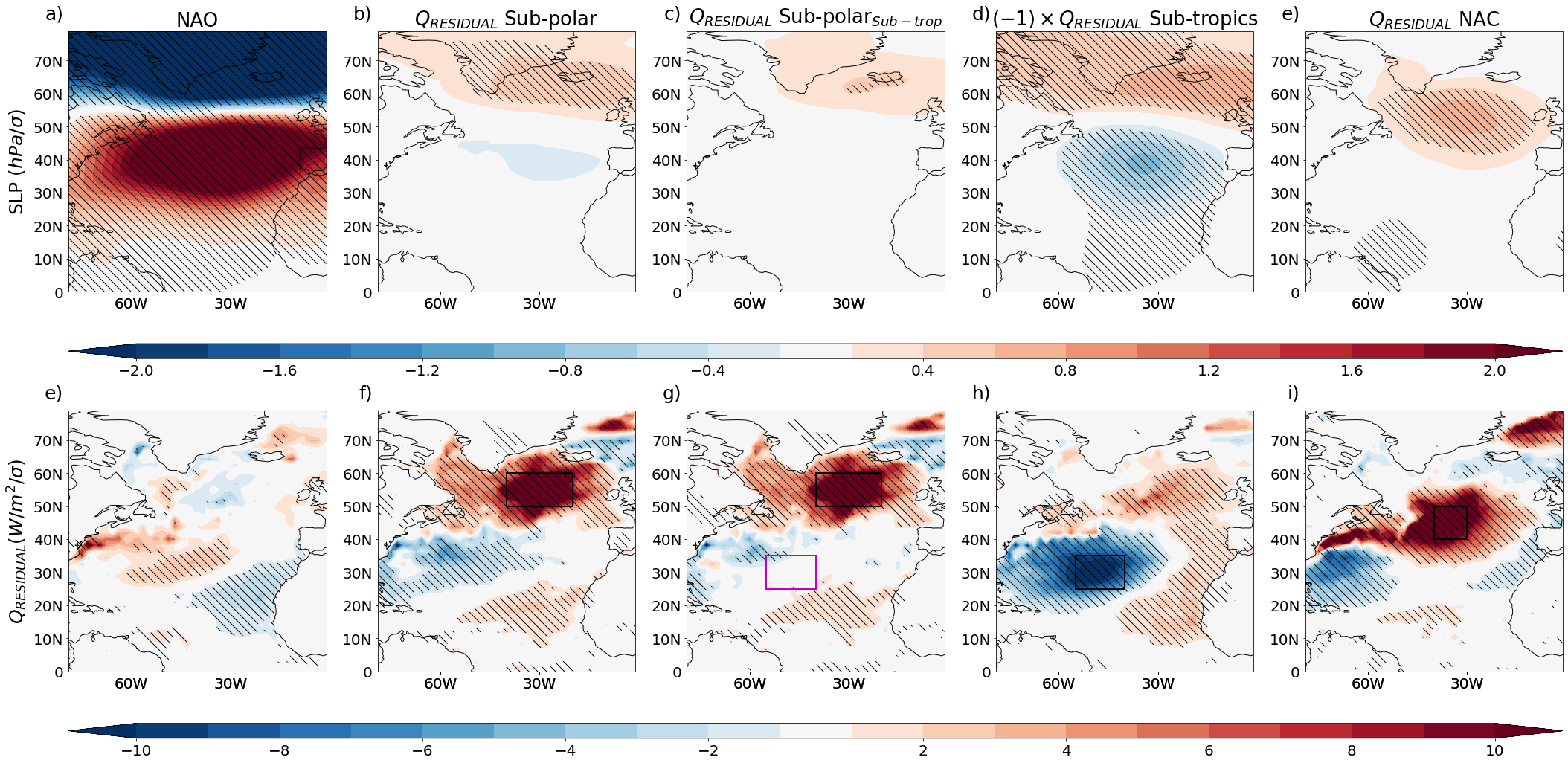}    
    \caption{Regression analysis of the connection between SLP and $Q_{RESIDUAL}$ variability in the piControl run. Regression of  a) SLP and e) $Q_{RESIDUAL}$ onto the NAO index. The other panels show regression of b-d) SLP and f-h) $Q_{RESIDUAL}$ onto the mean $Q_{RESIDUAL}$ in boxes over the b,f) sub-polar (50$^{\circ}$N-60$^{\circ}$N,40$^{\circ}$W-20$^{\circ}$W) c,g) subtropical (25$^{\circ}$N-35$^{\circ}$N,55$^{\circ}$W-40$^{\circ}$W) and d,g) NAC (40$^{\circ}$N-50$^{\circ}$N, 40$^{\circ}$W-30$^{\circ}$W) regions. Each of these regions is indicated by black boxes in the respective panels. The magenta box in g) indicates the sub-tropical box which has been linearly removed from the sub-polar index. Hatching in a-h) indicates where regression coefficients are statistically significant, with p-values below 0.05, following a Student's t-test. }
\label{THF}
\end{figure} 

\subsection{Forcing by $Q_{RESIDUAL}$ in different regions}
To further interrogate the relative importance of the different centres of action, we regress $Q_{RESIDUAL}$ and SLP onto the NAO. The response to $Q_{RESIDUAL}$ EOF2 projects onto the negative phase of the NAO (figure \ref{EOFs}l), hence it is surprising that the NAO-$Q_{RESIDUAL}$ regression shows near zero regression coefficients over the eastern sub-polar North Atlantic (figure \ref{THF}f), as this location features prominently in $Q_{RESIDUAL}$ EOF2 (figure \ref{EOFs}j). Instead, the Gulf Stream and central to eastern sub-tropics show the largest $Q_{RESIDUAL}$-NAO relationship (figure \ref{THF}f). 

To further analyse the forcing of atmospheric circulation by $Q_{RESIDUAL}$ in different regions, we regress the SLP and $Q_{RESIDUAL}$ fields onto the mean $Q_{RESIDUAL}$ calculated in various latitude-longitude boxes. The mean $Q_{RESIDUAL}$ is calculated over boxes in the eastern sub-polar (50$^{\circ}$N-60$^{\circ}$N,40$^{\circ}$W-20$^{\circ}$W), sub-tropical (25$^{\circ}$N-35$^{\circ}$N,55$^{\circ}$W-40$^{\circ}$W) and NAC (40$^{\circ}$N-50$^{\circ}$N,40$^{\circ}$W-30$^{\circ}$W) regions. Regression with the eastern sub-polar region appears to suggest a weak high pressure response over Iceland from positive $Q_{RESIDUAL}$ in this region (figure \ref{THF}b). However, $Q_{RESIDUAL}$ in the eastern sub-polar region also shows a negative correlation with $Q_{RESIDUAL}$ in the subtropics (figure \ref{THF}g), leaving some ambiguity about which region is forcing the atmosphere. To better separate out the effects of forcing from these regions, we regress the sub-tropical box out of the sub-polar box time series'. Using this new sub-polar time series, the atmospheric circulation response is much weaker, suggesting that the eastern sub-polar region has little effect on atmospheric circulation (figure \ref{THF}c). Regression of SLP and $Q_{RESIDUAL}$ with the sub-tropical box in figure \ref{THF}d,i) confirms the importance of the western sub-tropical North Atlantic in driving a negative NAO response, in agreement with \citet[][their figure 1]{baker_linear_2019}. For completeness, high $Q_{RESIDUAL}$ along the NAC, as in $Q_{RESIDUAL}$ EOF1, forces a similar pattern to EOF1 (compare figure \ref{EOFs}k with figure \ref{THF}e,j), suggesting that the warm NAC is indeed forcing the ridge over the North-east Atlantic.


 
\begin{figure}
  \centering
    \includegraphics[width=\textwidth]{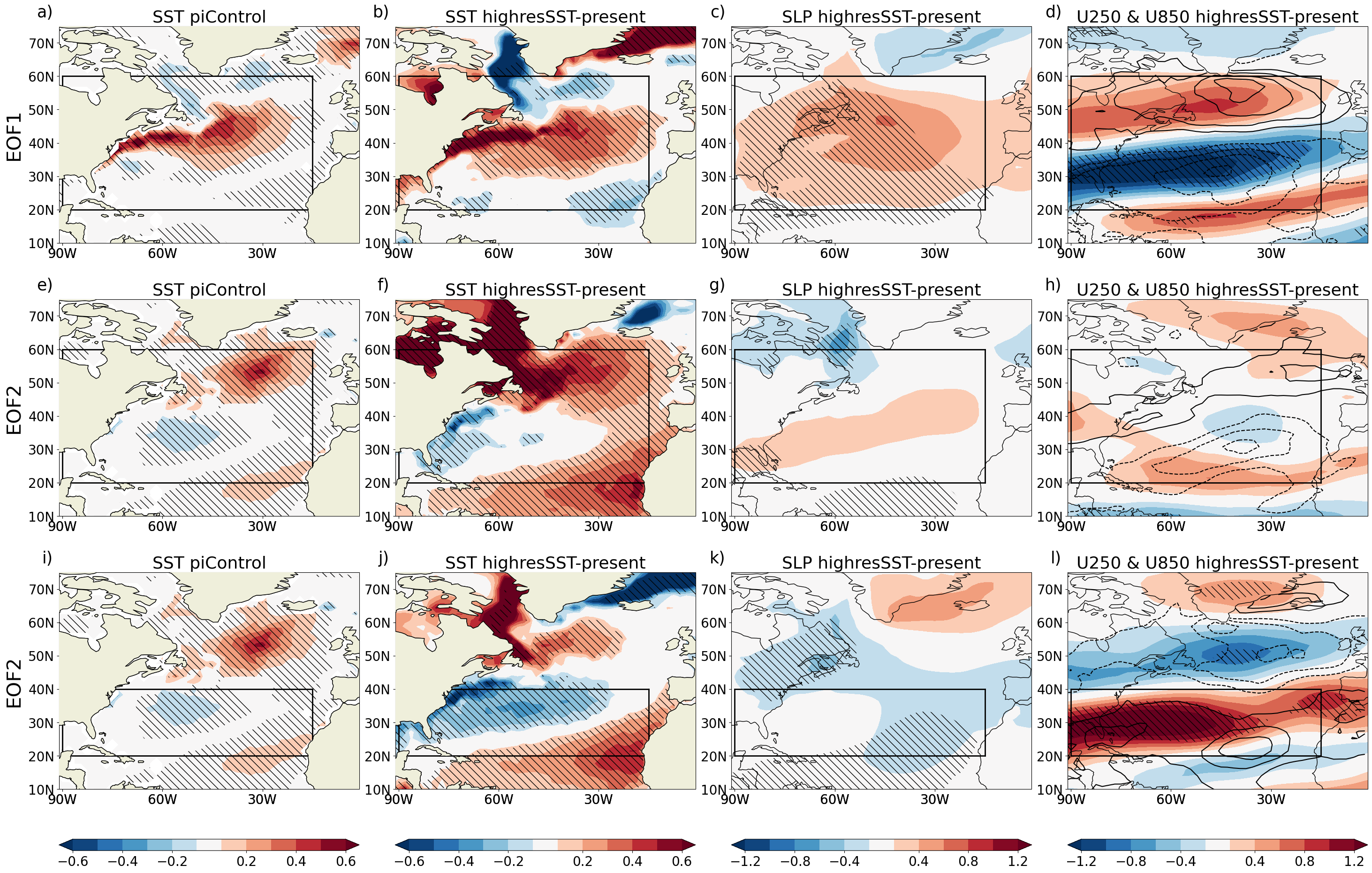}    
    \caption{Regressions of SSTs onto $Q_{RESIDUAL}$ EOFs 1 and 2 in the piControl run and regressions of SST, SLP and U850 onto an index of SST variability in highresSST-present experiments. The index is the projection of the SSTs in the highresSST-present onto SST patterns associated with $Q_{RESIDUAL}$ EOFs 1 and 2 in the piControl simulation. Hatching indicates where regression coefficients are statistically significant, with p-values below 0.05, following a Student's t-test. }
\label{amip}
\end{figure} 

\subsection{Atmosphere-only experiments}
Further evidence that the circulation pattern correlated with the $Q_{RESIDUAL}$ EOF1 is forced by the SSTs is provided by analysis of the highresSST-present atmosphere-only simulations. The highresSST-present simulations are run using the same model as the piControl simulations, but forced with observed SSTs and historical greenhouse gas and aerosol forcings spanning 1950-2014. The SSTs do not react to changes in circulation patterns and hence the direction of causality is clear. There are three ensemble members of highresSST-present and we take the ensemble average as the SSTs are the same in each case. We project the SST patterns associated with $Q_{RESIDUAL}$ EOF1 and EOF2 onto SSTs in the highresSST-present simulations, first over a box spanning (20$^{\circ}$N-60$^{\circ}$N,90$^{\circ}$W-15$^{\circ}$W), which is shown in figure \ref{amip}a-h). We then regress SSTs, SLP and zonal wind onto the resulting time series. It should be noted that the piControl and highresSST-present SST patterns are not identical (figure \ref{amip}a,b and figure \ref{amip}e,f) as the highresSST-present simulations are forced by observed SSTs, which will have different modes of SST variability. For example, the Labrador Sea is substantially more prominent in the SST pattern in highresSST-present for $Q_{RESIDUAL}$ EOF2, while the region of cool sub-tropical SSTs is shifted westwards relative to the piControl (figure \ref{amip}e,f).

Nevertheless, atmospheric circulation appears to respond similarly to the $Q_{RESIDUAL}$ EOF1-like pattern in the highresSST-present as in the piControl. The highresSST-present simulations show high pressure over the North-West Atlantic and a northward shifted jet (figure \ref{amip}c,d), similar to the piControl (figures \ref{EOFs}k and \ref{response}b). In contrast, the $Q_{RESIDUAL}$ EOF2-like pattern appears to show a very weak response in highresSST-present, and opposite in sign to the piControl (figure \ref{amip}g,h and figures \ref{EOFs}l and \ref{response}f). However, given the prominence of the sub-polar region in figure \ref{amip}f) and lack of a circulation response to sub-polar SSTs in the piControl run, we instead project the $Q_{RESIDUAL}$ EOF2 SST pattern in the piControl onto highresSST-present SSTs over only latitudes 20$^{\circ}$N-40$^{\circ}$N (box in figure \ref{amip}i-l) to emphasise the sub-tropical portion. In this case, there is a weak negative NAO-like dipole pattern in SLP  (figure \ref{amip}k) and southward shifted jet (figure \ref{amip}l), in better agreement with the piControl. This further emphasises the importance of the sub-tropics for forcing an atmospheric circulation response to North Atlantic SST variability. 

\begin{figure}
  \centering
    \includegraphics[width=\textwidth]{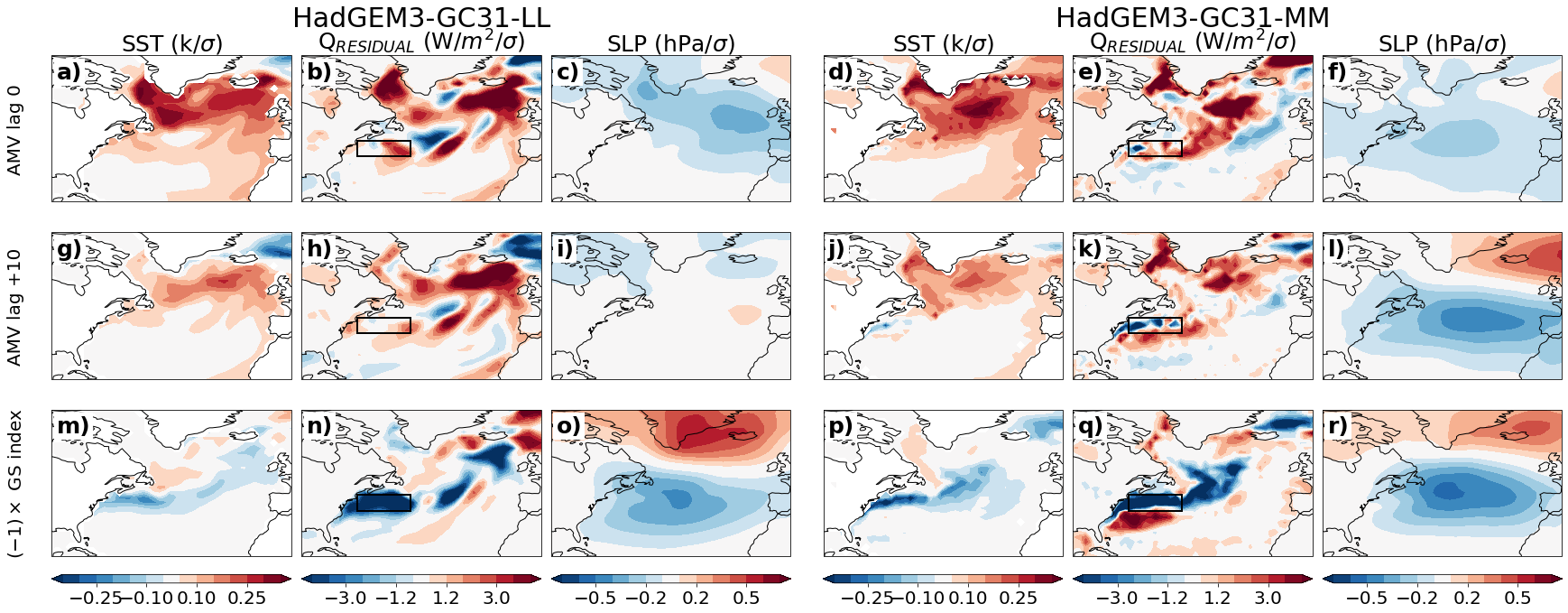}    
    \caption{Variables regressed onto the AMV index in the (left) LL and (right) MM versions of HadGEM3-GC3.1. Shown are regressions of 15-year running mean a,d,g,j,m,p) SST, b,e,h,k,n,q) $Q_{RESIDUAL}$ and c,i,o,f,l,r) SLP onto a-f) AMV at lag 0, g-l) the AMV index but for the AMV leading by 10 years and m-r) a Gulf Stream index (see text). The box used to define the Gulf Stream index is shown in panels with $Q_{RESIDUAL}$ regressions.}
\label{AMV}
\end{figure} 

\section{Analysing AMV forcing of the atmosphere at different resolutions}
\label{Resolution}

As a final application of the dynamical decomposition method, we analyse the response of HadGEM3-GC3.1 to AMV with differing horizontal resolutions. \citet{lai_mechanisms_2022} showed that the details of the mechanisms driving multi-decadal variability differ between the MM and LL versions of the model, even though the SST patterns look relatively similar. Specifically, the atmosphere has a key role in forcing sub-surface density and AMOC changes which contribute to AMV in the MM version, whereas AMV is more ocean-driven in the LL version \citep{lai_mechanisms_2022}. A key difference appears to be that sub-surface density anomalies, associated with AMV, propagate southward into the Gulf Stream extension in MM. In contrast, LL sees persistent density anomalies largely confined to the sub-polar region. Interestingly, \citet{lai_mechanisms_2022} found that the MM version also shows a negative NAO-like response following the AMV peak (figure \ref{AMV}l), whereas the LL version shows no response (figure \ref{AMV}i). We investigate this latter result now using $Q_{RESIDUAL}$. 

At the peak of AMV, both LL and MM versions have the characteristic horseshoe-like SST pattern (figure \ref{AMV}a,d) and are associated with negative SLP anomalies in the North Atlantic (figure \ref{AMV}c,f). The negative SLP anomalies are likely associated with the circulation pattern driving the AMV (whether directly via surface heat fluxes or indirectly via changes to AMOC strength). Ten years after the AMV peak, the LL and MM SST patterns are relatively similar with one difference being in the Gulf Stream region (figure \ref{AMV}g,j), where MM shows a weak negative SST anomaly. Given their similarity, what is the cause of the difference in circulation responses to AMV? 

Regressions with $Q_{RESIDUAL}$ reveal more of a difference between the LL and MM simulations at lag 10, again particularly in the Gulf Stream region (figure \ref{AMV}h,k). In particular, the MM run shows a negative $Q_{RESIDUAL}$ anomaly at lag 10 in the boxed region (figure \ref{AMV}k), whereas the LL simulation shows no substantial anomaly (figure \ref{AMV}h). To provide further evidence that the Gulf Stream anomalies are the reason for the difference in responses to AMV between the different versions, we compute a Gulf Stream index as the mean $Q_{RESIDUAL}$ within the boxed region shown in figure \ref{AMV} (70$^{\circ}$W-50$^{\circ}$W,37$^{\circ}$N-43$^{\circ}$N) and regress the variables onto this index. We also multiply the Gulf Stream index by $-1$ such that $Q_{RESIDUAL}$ has the same sign as following a positive AMV. For both LL and MM versions, this produces an SLP pattern similar to the AMV lag 10 for MM, with a low pressure anomaly in the mid-latitudes and high pressure north-west of the UK (figure \ref{AMV}o,r). This is also akin to the observed response to Gulf Stream SST identified by \citet{wills_observed_2016}. This analysis suggests that rather than AMV primarily influencing atmospheric circulation via the sub-polar North Atlantic, that it is the Gulf Stream region which plays a leading role, at least in HadGEM3-GC3.1. Moreover, the differing atmospheric circulation responses to AMV in the different model versions stem from the fact that AMV-related sub-surface density anomalies propagate into the Gulf Stream region in MM, but not in LL.

\section{Discussion and conclusions}
\label{Conclusions}


This study has presented a method to separate the turbulent heat flux ($Q$) into a component directly related to atmospheric circulation and a residual component, $Q_{RESIDUAL}$, which is assumed to be primarily forced by ocean variability. The method uses a circulation analogues technique to quantify the circulation-related component and has been tested using the HadGEM3-GC3.1-MM pre-industrial (piControl) run for the wintertime North Atlantic. The leading modes of $Q_{RESIDUAL}$ show substantial low frequency variability and the peak of $Q_{RESIDUAL}$ EOF1 is closely linked with a strengthening of the AMOC. The modes are characterised by a warming of the atmosphere along the Gulf Stream and North Atlantic Current (NAC) for the EOF1; and a dipole of $Q$ anomalies with cooling of the atmosphere in the western subtropical North Atlantic and warming in the eastern sub-polar region for EOF2. The first and second EOFs also drive equivalent barotropic atmospheric circulation responses in the form of Atlantic ridge and negative NAO patterns, respectively.

A key result of this work was that the Gulf Stream and western sub-tropical North Atlantic play a much larger role in driving the  atmospheric circulation response to SST variability in HadGEM3-GC3.1-MM relative to the sub-polar region. This result was evidenced by analysis of $Q_{RESIDUAL}$ in the piControl simulations and of atmosphere-only simulations. This also explains a result of \citet{lai_mechanisms_2022}, who found that the medium (MM) resolution version of HadGEM3-GC3.1 showed a robust negative-North Atlantic Oscillation (NAO) response to Atlantic Multidecadal Variability (AMV), in contrast to the lower (LL) resolution version of the model. They showed that sub-surface density anomalies propagated considerably further south into the Gulf Stream region along the western boundary of the North Atlantic following the peak of an AMV event in the MM version than in the LL version. These anomalies likely contributed to the $Q_{RESIDUAL}$ patterns seen in MM, which forced an atmospheric circulation response. 

Although beyond the scope of this paper, it would be of interest to apply the $Q$ dynamical decomposition method to other regions. For example, Indian Ocean sea surface temperatures (SSTs) are strongly affected by atmospheric variability, but also can force atmospheric circulation anomalies. Hence, the dynamical decomposition could provide a useful diagnostic for separating the atmospheric and ocean-driven $Q$ patterns in this region. Other regions which could be of interest include the tropical Atlantic and North Pacific. 

It should be noted that the dynamical decomposition method is primarily a diagnostic tool, for separating atmospheric and ocean-driven components of $Q$. That is to say, the atmosphere responds to the total $Q$ and not $Q_{RESIDUAL}$. \citet{oreilly_challenges_2023} identified differences in the sign of $Q$ anomalies in free-running, coupled model experiments compared with idealised pacemaker experiments. Specifically, restoring tropical Atlantic SSTs towards particular patterns often results in positive $Q$, though in coupled runs, warm SSTs in this location are usually linked to negative $Q$. The dynamical decomposition method could be of particular use in comparing $Q$ in free-running models to that in pacemaker experiments and thus establishing whether the SST-restoring primarily occurs through atmospheric or oceanic adjustment. 

The dynamical decomposition method requires a few choices of parameters, namely the number of samples to fit for each year ($N_s$), the number of similar years to randomly select from ($N_a$) and the number of times that this process is repeated ($N_r$). However, the results were not found to be overly sensitive to these parameters and an analysis of variations in these parameters is given in appendix A. 

A larger uncertainty concerns the length of the time series used to estimate the circulation response to SST variability and this is further discussed in appendix B. Briefly, shorter historical simulations, forced by variable greenhouse gases and aerosols, using the same model as for the piControl simulation, produce similar $Q_{RESIDUAL}$ patterns to the piControl. However, the atmospheric circulation response shows considerably more uncertainty in the historical runs. Mid-latitude SST variability which appreciably affects atmospheric circulation, primarily occurs on decadal to multidecadal timescales as the forcing itself is weak and must therefore be persistent. Hence, even for a 100-year dataset, this may be inadequately sampled. This would therefore suggest that caution should be exercised in studying the circulation response in shorter model runs or reanalyses datasets. Nevertheless, there is significant evidence that the response to mid-latitude SSTs is underestimated in models compared to the real world \citep{eade_seasonal--decadal_2014,scaife_signal--noise_2018,smith_north_2020}, and that models underestimate the true multidecadal variability of North Atlantic circulation \citep{simpson_modeled_2018,oreilly_projections_2021}. Therefore, 100-year long reanalysis and other observation-based datasets may still show an appreciable signal. We aim to examine this in a future paper.

\clearpage
\acknowledgments
MP, JR and TW were funded through the Natural Environment Research Council (NERC) project WISHBONE (NE/T013516/1) and COR was funded by a Royal Society University Research Fellowship. JR was also funded through the NERC project CANARI (NE/W004984/1). The authors thank the Met Office for making data produced using the HadGEM3-GC3.1 model available and thank Dr Michael Lai for sharing the AMOC stream function time series'. The authors also thank Professor Richard Greatbatch for comments on a draft of this manuscript and thank colleagues at Oxford and Reading for helpful discussions on this work. 


%
%
\datastatement
The HadGEM3-GC3.1 data used in this paper were downloaded from the Earth System Grid Federation website (e.g. https://esgf-index1.ceda.ac.uk/projects/esgf-ceda/). ERA5 data were downloaded from the Climate Data Store website (https://cds.climate.copernicus.eu/cdsapp\#!/dataset/reanalysis-era5-pressure-levels-monthly-means?tab=overview) and HadISST2 from the Met Office website (https://www.metoffice.gov.uk/hadobs/hadisst2/data/download.html).  

%



\appendix[A]
\appendixtitle{Sensitivity of circulation analogues to choice of $N_r$, $N_s$ and $N_a$}
\label{Parameters}

\begin{figure}
  \centering
    \includegraphics[width=\textwidth]{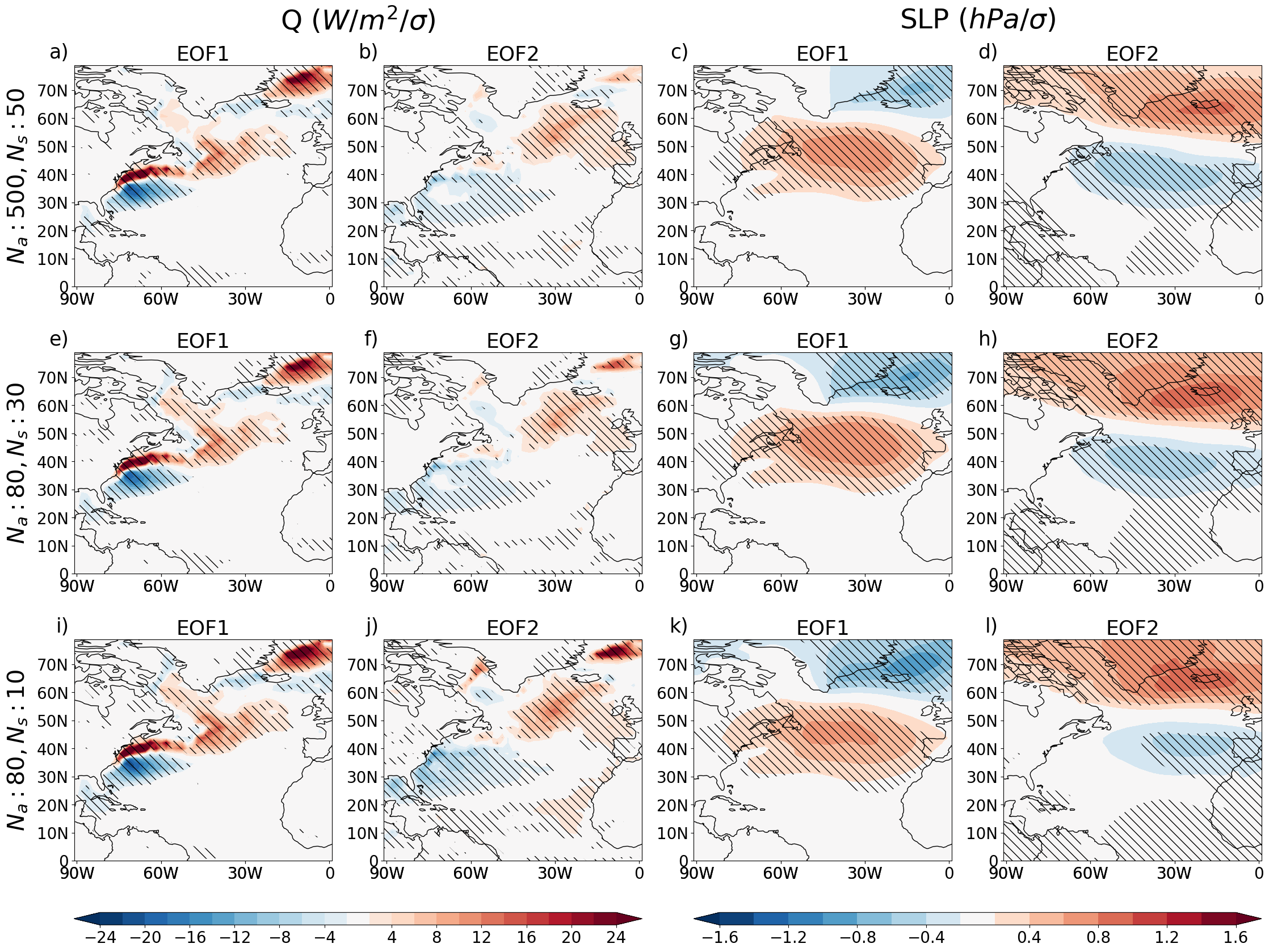}    
    \caption{As in figure \ref{EOFs} but for $Q_{RESIDUAL}$ only and for different values of $N_s$ and $N_a$. Shown are results for  a-d) $N_s=50$, $N_a=500$, e-h) $N_s=30$, $N_a=80$ and i-l) $N_s=10$, $N_a=80$.}
\label{N_s}
\end{figure} 

We test for sensitivity of the $Q$ dynamical decomposition to variations in a number of parameters. We alter the number of closest years ($N_a$) and the number of years sub-sampled ($N_s$) each time and calculate the first EOFs. The $Q$ dynamical decomposition procedure only sub-samples from the nearest $N_a=80$ years, however relaxing this constraint and sub-sampling from all 500 years makes no different to the structure of the leading $Q_{RESIDUAL}$ modes (compare figure \ref{N_s}a-d with figure \ref{EOFs}i-l). The modes also do not appear to be sensitive to the choice of $N_s$, with both $N_s=30$ (figure \ref{N_s}e-h) and $N_s=10$ (figure \ref{N_s}i-l) giving similar solutions to $N_s=50$ (figure \ref{EOFs}i-l).   

\begin{figure}
  \centering
    \includegraphics[width=0.5\textwidth]{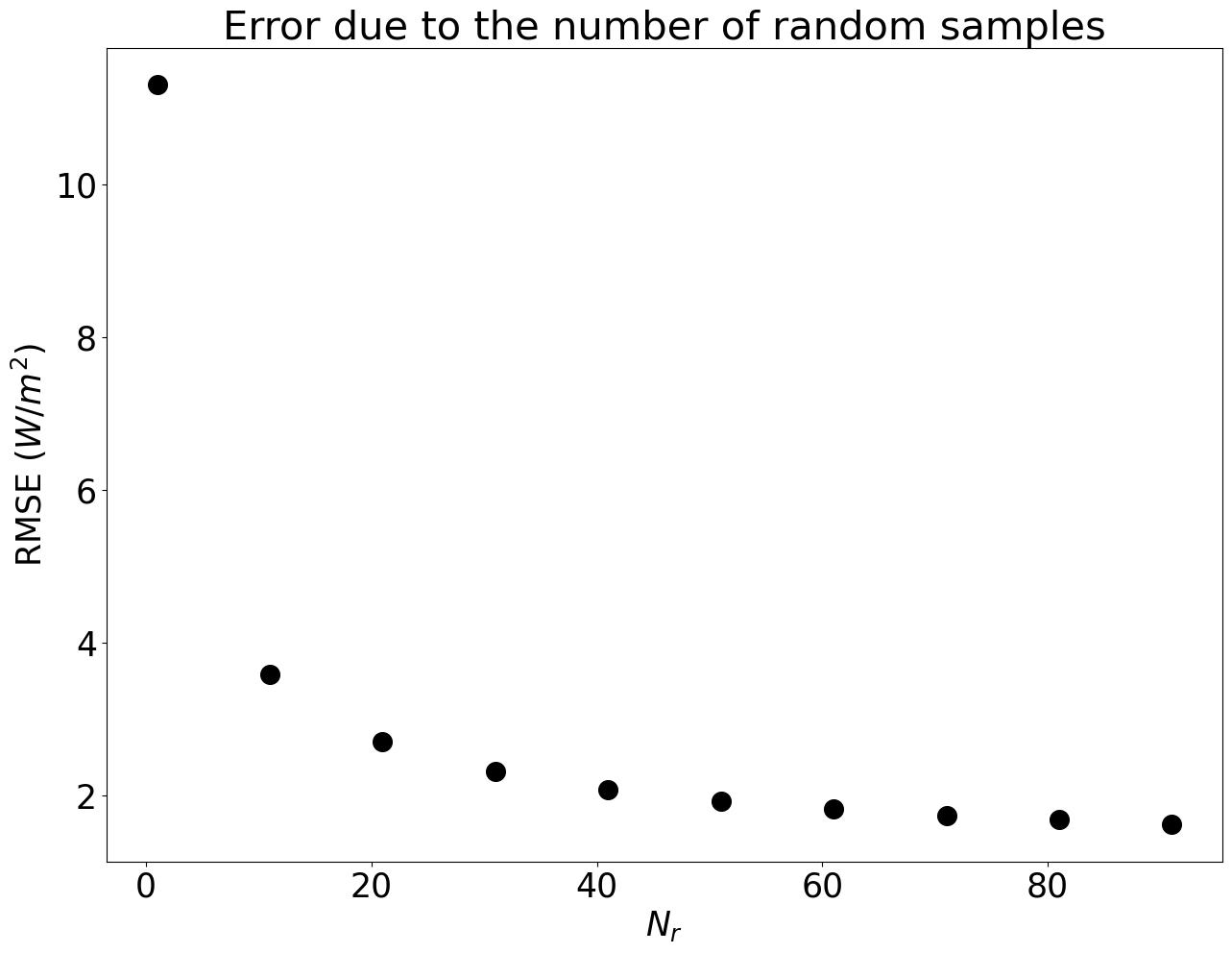}    
    \caption{The root mean square error in $Q_{RESIDUAL}$ when compared to the original $Q$ dynamical decomposition with $N_r=100$ for different values of $N_r$.}
\label{N_r}
\end{figure} 

Finally, the root mean square error (RMSE) in $Q_{RESIDUAL}$ when compared to the original $Q$ dynamical decomposition with the number of repeats, $N_r$, set to 100, levels off from around $N_r=30$ (figure \ref{N_r}). The results are therefore not sensitive to variations in $N_r$, apart from for very low values. The lack of sensitivity to the choice of $N_r$, $N_s$ and $N_a$ is consistent with sensitivity tests by \citet{deser_forced_2016}.

\appendix[B] 
\appendixtitle{Sensitivity to period length and external forcing}
\label{Sensitivity}

This study has presented the results of applying the circulation analogues method to a piControl run with a long time series and no externally-forced variability. It would however be desirable to apply this to historical simulations or reanalysis datasets which are shorter and are subject to variations in greenhouse gases and aerosols. This appendix therefore investigates the sensitivity of the method to the length of the period and presence of external forcing. To test the sensitivity to the length of the time period, we perform the dynamical decomposition using five 100-year and three 165-year, non-overlapping subsets of the piControl run. We also apply the dynamical decomposition to a four-member ensemble of runs with observed external forcings spanning 1850-2014 (historical), which have been created using the same model. In each case, the same values of $N_s$, $N_r$ and $N_a$ are used as for the original piControl run. For the historical simulations, the influence of external forcing is removed by linearly regressing out the global-mean SST from all variables before performing the dynamical decomposition.

\begin{figure}
  \centering
    \includegraphics[width=\textwidth]{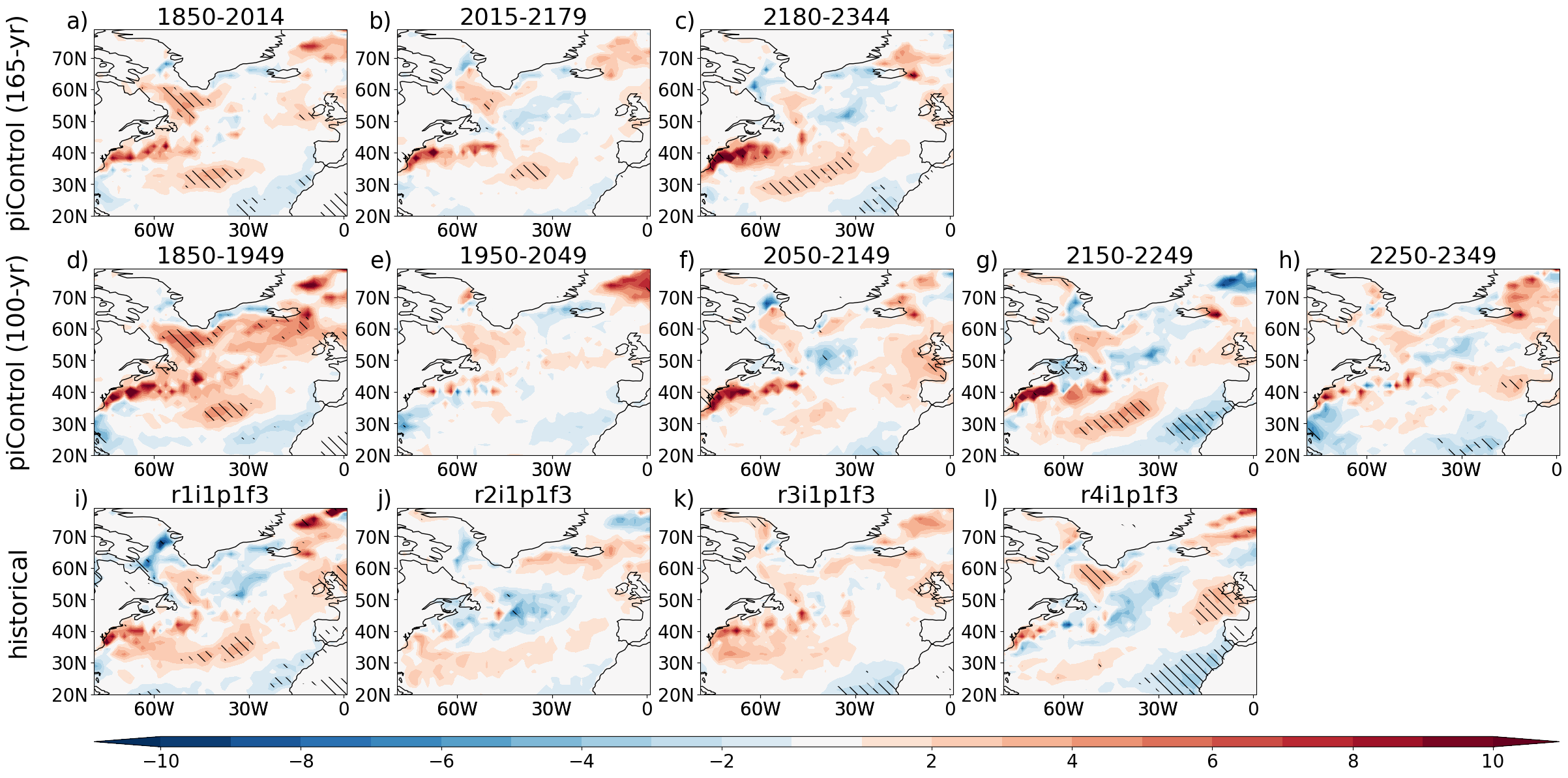}    
    \caption{Uncertainty associated with calculating SST forcing of the NAO using shorter periods of data. Results of regressing $Q_{RESIDUAL}$ onto the NAO are shown for a-c) the piControl run split into three 165-year periods, d-h) the piControl run split into five 100-year periods and i-l) four historical runs using the same model. Hatching indicates where regression coefficients are statistically significant, with p-values below 0.05, following a Student's t-test. Units are $W/m^2/\sigma$.}
\label{sensitivity}
\end{figure} 

Figure \ref{sensitivity} shows the sensitivity of the NAO to forcing by anomalous $Q_{RESIDUAL}$, as in figure \ref{THF}f), but calculated using subsets of the piControl run and for historical simulations. The 165-year subsets of the piControl are similar to one another and to the results from the full 500-year period. For instance, all show the positive phase of the NAO to be linked to anomalously high $Q_{RESIDUAL}$ over the Gulf Stream, a centre of action at 40$^{\circ}$W,30$^{\circ}$N and over the Labrador Sea (figure \ref{sensitivity}a-c). The 100-year subsets are also broadly similar to one another, but the strength of the different centres of action varies considerably between periods. The 1850-1949 and 2150-2249 periods show a relatively strong forcing of the NAO from the region around 30$^{\circ}$N,40$^{\circ}$W, but this correspondence is much weaker and not significant in the other periods (figure \ref{sensitivity}d-h). Similarly, the historical simulations show roughly similar patterns overall, again with positive NAO-$Q_{RESIDUAL}$ correlations around 30$^{\circ}$N,40$^{\circ}$W and over the Labrador Sea, but with the strength of the connection varying considerably between runs. 

\begin{figure}
  \centering
    \includegraphics[width=0.8\textwidth]{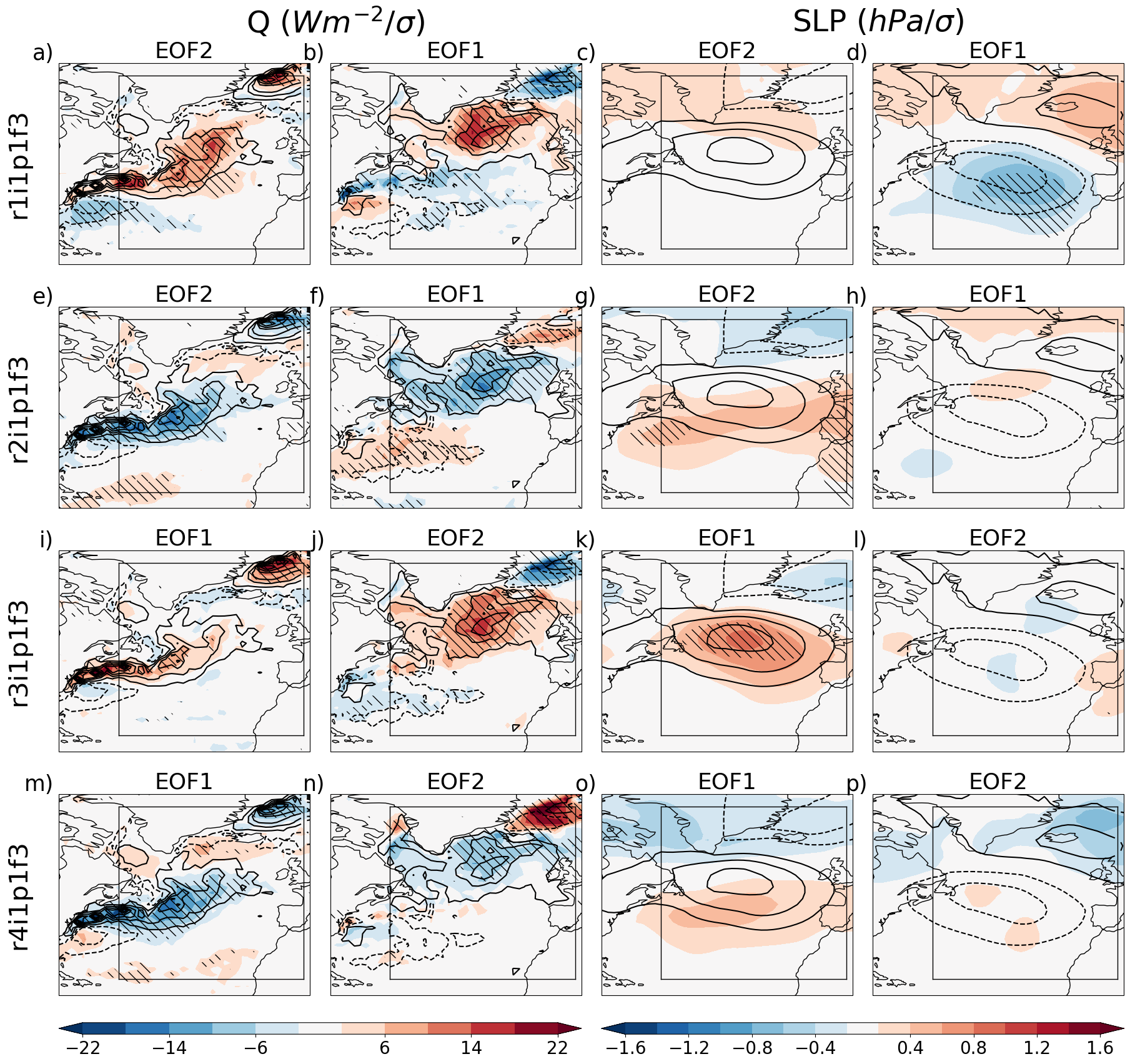}    
    \caption{Regressions of $Q$ and $SLP$ onto PCs associated with $Q_{RESIDUAL}$ EOF1 and EOF2, for the four historical ensemble members, is shown by colours. Results from the piControl simulations are shown by unfilled contours with the same contour interval. Hatching indicates where regression coefficients are statistically significant, with p-values below 0.05, following a Student's t-test.}
\label{historical}
\end{figure} 

\clearpage

Performing the EOF analysis of $Q_{RESIDUAL}$ as before, the historical simulations each show similar dominant patterns of $Q_{RESIDUAL}$ variability to the piControl, though for ensemble members r1i1p1f3 and r2i1p1f3 the order of the first two EOFs is flipped. This suggests that external forcing and the shorter length of the simulations does not prevent identification of the leading modes by which SST impacts the atmosphere. Furthermore, this provides addional evidence that the dynamical decomposition could usefully be applied to reanalysis datasets, which have the added complexity of external forcing and are considerably shorter than a typical piControl run. However, only r1i1p1f3 EOF1, r3i1p1f3 EOF1 and r4i1p1f3 EOF2 show similar SLP responses to the piControl and indeed r2i1p1f3 EOF2 and r4i1p1f3 EOF1 show opposite responses, suggesting that the SLP response to the $Q_{RESIDUAL}$ modes is either non-stationary or overwhelmed by internal variability for the shorter simulations. 



%



\bibliographystyle{ametsocV6}
\bibliography{templateV6.1}

\end{document}